\documentclass[12pt]{article}
\pdfoutput=1
\usepackage[nosort]{cite}
\usepackage[normalem]{ulem}

\usepackage[dvipsnames]{xcolor}
\usepackage{epsfig}
\usepackage{amsfonts}
\usepackage{amscd}
\usepackage{latexsym}
\usepackage{amsmath,amssymb}
\usepackage{verbatim}
\usepackage{setspace}
\usepackage{color}
\usepackage{fancyhdr}
\usepackage{cite}
\usepackage{hyperref}
\usepackage{tikz}
\usepackage{slashed}
\usepackage{multirow}
\usetikzlibrary{calc}
\usetikzlibrary{topaths}
\usetikzlibrary{decorations}
\usetikzlibrary{decorations.pathmorphing}
\usetikzlibrary{arrows,decorations.markings,cd}
\usetikzlibrary{calc,arrows,cd,decorations.markings,snakes}
\tikzset{
->-/.style args={#1rotate#2}{decoration={markings, mark=at position #1 with {\arrow[scale=1.5,rotate = #2 ]{stealth}}}, postaction={decorate}}
}
\usetikzlibrary{shapes.geometric}
\usetikzlibrary{knots}
\usepackage{tikz-3dplot}

\usepackage{draft}
\usepackage{hyperref}
\usepackage{graphicx,color,subfig}
\usepackage{cite}
\usepackage{mciteplus}
\usepackage{skak}
\usepackage{bbm}

\numberwithin{equation}{section}

\def\ee{\mathrm{e}}
\def\mm{\mathrm{m}}
\def\oo{\mathrm{o}}

\def\bZ{\mathbb{Z}}

\def\SO{\mathrm{SO}}
\def\SU{\mathrm{SU}}
\def\U{\mathrm{U}}
\def\O{\mathrm{O}}

\def\c{\mathrm{c}}
\def\IR{\mathrm{IR}}
\def\UV{\mathrm{UV}}

\def\({\left(}
\def\){\right)}

\begin{document}

\begin{titlepage}

\title{Symmetry Transmutation and Anomaly Matching}

\author{Nathan Seiberg and Sahand Seifnashri}

\address{School of Natural Sciences, Institute for Advanced Study, Princeton, NJ}

\abstract
\noindent
We explore a situation where a global symmetry of the ultraviolet (UV) theory does not act faithfully on the local infrared (IR) degrees of freedom, but instead acts effectively as a higher-form symmetry.  We refer to this phenomenon as \emph{symmetry transmutation}, where the UV symmetry is ``transmuted'' into a higher-form symmetry in the IR.  Notably, unlike emergent (accidental) symmetries, which are approximate, these symmetries are exact. We illustrate the ubiquity of this phenomenon in various continuum and lattice systems and provide examples where the 't Hooft anomalies of the UV symmetry are matched by those of the new higher-form symmetry in the IR.  We also show that in certain phases and for certain energies, the UV baryon-number symmetry of one-flavor QCD is transmuted into a discrete one-form global symmetry.  Finally, we compare our symmetry transmutation to the well-known phenomenon of symmetry fractionalization.
\end{titlepage}

\tableofcontents

\section{Introduction}

Symmetries play a fundamental role in many branches of science, including chemistry, condensed matter physics, and high-energy physics, serving as a key guiding principle in their analysis. In particular, understanding how the symmetries of the short-distance (UV) theory are realized at long distances (IR) leads to powerful constraints on the latter.  The symmetries and their manifestations organize the possible phases of the system and the phase transitions between them.

Following \cite{Gaiotto:2014kfa}, recently, the concept of global symmetries has been generalized in several directions. (For reviews, see \cite{McGreevy:2022oyu,Cordova:2022ruw,Schafer-Nameki:2023jdn,Shao:2023gho}.) Can one ignore these generalized symmetries and focus only on conventional (zero-form) symmetries? In this note, we will see that IR \emph{generalized symmetries} that are not present in the microscopic theory can have profound implications on the consistency of the system, and they should not be overlooked.  For example, such IR higher-form symmetries are essential in understanding the 't Hooft anomaly matching between the UV and the IR.

It is well known that even when a system has only conventional zero-form symmetries at the microscopic level, higher-form symmetries can emerge at long distances. In this work, we study a different scenario, where higher-form symmetries at long distances originate directly from the zero-form symmetries of the microscopic theory rather than being emergent or accidental.

Specifically, we explore the phenomenon of \emph{symmetry transmutation}, where the global symmetry of the UV theory does not act faithfully on the low-energy degrees of freedom, but instead, it acts in the IR effectively as a higher-form symmetry. We say the UV low-form symmetry is \emph{transmuted} in the IR into a higher-form symmetry.\footnote{The use of the term ``transmute'' for this phenomenon is particularly appropriate because, according to Merriam-Webster, to transmute means ``to change or alter in form, appearance, or nature and especially to a higher form.''}

It is important to point out that many of our examples are not new, and the matching of their symmetries and anomalies is also understood.  (We will refer to the relevant papers when we discuss the examples.)  The novelty in our presentation is in the unified treatment of many examples as a single phenomenon and in using this understanding to extend it to new systems.

\subsection{Matching symmetries and anomalies between the UV and the IR}

It is often stated that there is a homomorphism between the symmetry group of the UV theory, $G_\UV$, and the symmetry group of the IR theory, $G_\IR$:
\ie
 	\varphi: G_\UV  \to G_\IR \,. \label{homomorphism}
\fe
We take $G_\UV$ and $G_\IR$ to be faithful symmetries of the UV and IR theory, respectively. See Section \ref{necessary conditions}, for a discussion of faithfulness in this context.
This means that every symmetry operator in the UV maps to a symmetry operator in the IR. However, there are two important details to consider:
\begin{itemize}
\item First, the IR theory might have emergent (accidental) symmetries that do not originate from the UV.  These are approximate IR symmetries that are not in the image of the map $\varphi$, $\mathrm{Im}(\varphi)$.
    Emergent symmetries are in the co-kernel of $\varphi$.
\item More importantly for us, the homomorphism $\varphi$ might have a nontrivial kernel. This means that some UV symmetries act trivially in the IR.  Equivalently, $G_\UV$ might not act faithfully in the IR, and then
    \ie\label{GUVquotient}
    \mathrm{Im}(\varphi)  \cong {G_\UV\over \mathrm{Ker}(\varphi)}\,,
    \fe
    where $\mathrm{Ker}(\varphi)$ is the kernel of $\varphi$, which is a normal subgroup of $G_\UV$.
\end{itemize}

We will soon see that this standard picture is incomplete.

It is common to couple these symmetries to background gauge fields.  Every UV background gauge field $A_\UV $ of $G_\UV $ is then mapped to an IR background $A_\IR $ of $G_\IR $:
\ie
A_\IR  = \Phi (A_\UV )\,. \label{ordinary.map.Phi}
\fe
The map $\Phi$ is constructed from $\varphi$, and they carry the same information. Knowing the map $\Phi$ is essential for matching the 't Hooft anomalies of the UV and IR symmetries, in the sense that $\alpha_\UV [A_\UV ] = \alpha_\IR [\Phi(A_\UV )]$, where $\alpha[A]$ denotes the anomaly theory.

Just as for the homomorphism of the symmetry groups in \eqref{homomorphism}, the homomorphism of background gauge fields in \eqref{ordinary.map.Phi} has similar details.
\begin{itemize}
\item First, when the map $\varphi$ is not onto, the map $\Phi$ is also not onto. The map $\Phi$ does not map to gauge fields for the emergent symmetries that are not in the image of $\varphi$. Anomalies in such emergent symmetries do not have to be matched by the UV anomalies.
\item Even if the map $\varphi$ is onto (i.e., $ \mathrm{Im}(\varphi)=G_\IR$), the map of the gauge fields $\Phi$ might not be onto.  In particular,  the quotient by $\mathrm{Ker}(\varphi)$ in \eqref{GUVquotient} could lead to gauge fields of $ G_\IR$ that cannot be lifted to gauge fields of $G_\UV$.  Consequently, the symmetry of the IR, $G_\IR$, might have 't Hooft anomalies, which do not reflect any anomaly in $G_\UV$.  In \cite{Metlitski:2017fmd, Thorngren:2020wet}, such anomalies were referred to as \emph{emergent anomalies}.  Notably, this is distinct from anomalies in emergent symmetries.
\item Finally, most important for us, is a situation where the kernel of the map $\varphi$ \eqref{homomorphism} has an 't Hooft anomaly in the UV theory.  In that case, it seems that it cannot be matched by an anomaly in the IR.   Naively, such a situation is excluded by the 't Hooft anomaly matching conditions.  Instead, we will see that the anomalies can still be matched, provided the picture with the homomorphism \eqref{homomorphism} is extended.
\end{itemize}
Note that in the first and second bullet points above,  the IR theory has more anomalies than the UV theory.  Yet, this is consistent with the 't Hooft anomaly matching.  On the other hand, in the third bullet point,  the UV theory seems to have more anomalies than the IR theory, and an extension of the homomorphism picture \eqref{homomorphism} is needed.

In the next subsection, we will explain this extension.  We will see that the IR theory has a new higher-form symmetry and its anomalies are crucial in matching the UV anomalies that cannot be matched by the zero-form symmetry $G_\IR$.

\subsection{Symmetry transmutation}

The main point of this paper is to generalize the map \eqref{ordinary.map.Phi}.
Originally, this map relates background gauge fields for UV zero-form symmetries to IR zero-form symmetries. We extend it to background gauge fields for generalized symmetries, including higher-form symmetries, higher-groups, etc. We will continue to denote this generalized map by $\Phi$.

Below, we will denote by $G_\UV$ and $G_\IR$ the symmetries in the UV and the IR, respectively.  These can include various generalized symmetries.

First, consider a simple case where the UV theory has only zero-form symmetries with gauge fields $A_\UV$, and the IR theory has only a one-form symmetry with gauge field $B_\IR$.  (We will discuss more general cases below.)  In this case, we say that an ordinary 0-form symmetry of the UV  is \emph{transmuted} into a higher-form symmetry in the IR.  In this simple case, the generalization of the map \eqref{ordinary.map.Phi} becomes
\ie
	B_\IR  = \Phi(A_\UV ) \,. \label{map.Phi}
\fe
This map between the background gauge fields provides a formula for the IR higher-form symmetry operators in terms of the UV symmetry operators and \emph{defects}. Geometrically, this means that the IR symmetry operators and defects are constructed using the UV symmetry operators and defects.  See Section \ref{sec:geometric} for details.

We emphasize that examples of maps like \eqref{map.Phi} are standard and have appeared in numerous papers, including \cite{Kapustin:2013uxa,Benini:2018reh,Cordova:2019bsd,Delmastro:2022pfo, Brennan:2022tyl,Barkeshli:2022edm,Antinucci:2024bcm, Dumitrescu:2024jko}. The geometric version of this map, in terms of symmetry defects, has been discussed in \cite{Barkeshli:2014cna}.  We will interpret the map \eqref{map.Phi} as being associated with a new IR one-form global symmetry, which originates from a UV zero-form symmetry.  This interpretation as symmetry transmutation will allow us to organize many diverse examples in a unified fashion.

Importantly, the map \eqref{map.Phi} is not derived from a homomorphism as in \eqref{homomorphism}. How can we characterize the map $\Phi$ in terms of the (higher-form) symmetry \emph{groups} of the UV and IR? Let us focus, as in \eqref{map.Phi}, on the case of symmetry transmutation from a zero-form symmetry $G^{(0)}$ into a one-form symmetry $K^{(1)}$.\footnote{In our notation, $G^{(p)}$ denotes a $p$-form symmetry with symmetry group $G$. However, in order not to clutter the equations, we sometimes suppress the superscript $(0)$ for 0-form symmetries.} Below, we will show that in this case, the map $\Phi$ is characterized by a central extension of $G$ by $K$, or equivalently by the cohomology group
\ie
	 H^2(G, K)\,. \label{cohomology.group}
\fe
This data replaces the homomorphism \eqref{homomorphism}. We provide gauge theory examples that realize all such symmetry transmutation patterns. These correspond to gauge theories with gauge group $K$ and massive matter fields that transform under a central extension of $G$ by $K$.

Symmetry transmutation is closely related to symmetry fractionalization \cite{Jackiw:1975fn,Su:1981xk,Laughlin:1983fy,Senthil:1999czm,Essin:2013rca,Lu:2013jqa,Barkeshli:2014cna} (see also \cite{Chen:2014wse,Cheng:2015kce,Chen:2016fxq,Barkeshli:2019vtb,Manjunath:2020kne,Aasen:2021vva,Bulmash:2021hmb,Hsin:2024aqb, Kawagoe:2024urk, Rubio:2024tpv, Brennan:2025acl}), especially for the case of symmetry transmutation from 0-form symmetries into 1-form symmetries.\footnote{Symmetry transmutation of 0-form symmetries into 2-form symmetries is related to symmetry ``fractionalization'' for loop/string excitations; see, e.g., \cite{Cheng:2015kvt,Hsin:2019fhf,Ning:2019ffr,Chen:2020msl}.} In particular, the cohomology group \eqref{cohomology.group} is the same group that classifies symmetry fractionalization patterns for coupling a 2+1d bosonic TQFT with Abelian anyons described by $K$ to a zero-form $G$ symmetry \cite{Barkeshli:2014cna} (see \cite{Etingof:2009yvg,jones20233} for the underlying mathematical theory). The classic example is in the fractional quantum Hall effect, where the microscopic theory has a $G=\U(1)$ zero-form symmetry associated with charge conservation, and the macroscopic theory has anyons, including Abelian anyons associated with a one-form symmetry $K^{(1)}$. The microscopic U(1) symmetry, coupled to background gauge field $A_\UV$, acts effectively as a one-form symmetry at long distances through the formula
\ie
 	B_\IR = \mathrm{d} A_\UV\,,
\fe
where $B_\IR$ is a background gauge field for the one-form symmetry in the IR.\footnote{In many of the examples, the field $B$ in the left-hand-side is a gauge field of a discrete one-form global symmetry, denoted $\cal B$.  Then, this equation needs to be made more precise.  We will do that below. \label{BvscalB}} Specifically, the microscopic U(1) symmetry is transmuted into the one-form symmetry generated by the \emph{vison}. The TQFTs discussed in \cite{Vishwanath:2012tq,Metlitski:2013fal,Bonderson:2013pla,Chen:2013jha,Wang:2013uky,Wang:2013zja,Metlitski:2015eka, Mross:2014gla,Seiberg:2016rsg} can be interpreted as exhibiting such a symmetry transmutation associated with a UV anomalous time-reversal/U(1) symmetry in 2+1d.

Let us briefly discuss the relationship between symmetry \emph{fractionalization} and symmetry \emph{transmutation}. Symmetry fractionalization refers to the projective action of 0-form symmetries $G^{(0)}$ on anyons \cite{Barkeshli:2014cna}, or more generally, to non-trivial actions of $n$-group symmetries on $p$-dimensional operators with $p>n$. In symmetry fractionalization, both $G^{(0)}$ and $K^{(1)}$ can be present simultaneously. In contrast, symmetry transmutation describes a \emph{dynamical process} where $G^{(0)}$ appears only in the UV, while $K^{(1)}$ appears only in the IR as a result of transmutation. Symmetry transmutation concerns the distinction between emergent 1-form symmetries and exact 1-form symmetries that arise from a lower-form symmetry in the UV. Symmetry fractionalization is often discussed in the context of symmetry-enriched topological (SET) phases, where it constitutes the data used to classify the coupling of a 2+1d TQFT to a 0-form symmetry $G^{(0)}$ \cite{Barkeshli:2014cna}.  On the other hand, as we demonstrate here, symmetry transmutation occurs in any number of dimensions, both in gapped and gapless phases.

In summary:
\begin{itemize}
\item
Symmetry fractionalization starts with a given IR theory and its $K^{(1)}$ and analyzes the possible couplings of that theory to $G^{(0)}$.
\item
Symmetry transmutation starts with a given UV theory and describes how $G^{(0)}$ in the UV is realized in various phases of that theory in the IR, including how it can lead to $K^{(1)}$.
\end{itemize}

In systems where the UV global symmetry has an 't Hooft anomaly, it could be crucial to study symmetry transmutation to match the anomalies between the UV and the IR correctly.\footnote{We emphasize that symmetry transmutation is different from emergent one-form symmetries. In particular, the anomalies of emergent symmetries do not need to match and are not constrained by those in the UV. } We will discuss various gauge theory examples where the anomalies are matched through symmetry transmutation. In particular, we will discuss various phases of two-flavor scalar QED in 1+1d and 2+1d in Sections \ref{sec:2d.u1} and \ref{O3d3}. We will observe that these systems have different phases with different patterns of symmetry transmutation in each phase. Then, the symmetry transmutation allows us to match the UV and IR anomalies. Related anomaly matching has appeared in \cite{Cordova:2019bsd,Delmastro:2022pfo,Brennan:2022tyl,Antinucci:2024bcm, Dumitrescu:2024jko}.

We contrast symmetry transmutation with another phenomenon, where a higher-form symmetry leads to a \emph{lower-form} symmetry, see e.g., \cite{Heidenreich:2020pkc,Ferromagnets,Seiberg:2024yig}. For example, a theory in $d$-spatial dimensions with a $T^2$ target space, parameterized by $\phi^i\sim\phi^i+2\pi$ has two $U(1)$ ($d-1$)-form symmetries with currents $d\phi^i$ and a derived $U(1)$ ($d-2$)-form symmetry with current $d\phi^1 \wedge d\phi^2$. More generally, this corresponds to the higher-gauging construction of \cite{Roumpedakis:2022aik}, in which the symmetry defects/operators of the lower-form symmetry are constructed by gauging the higher-form symmetry on a submanifold inside spacetime. In this framework, only flat dynamical gauge fields are used, ensuring that the resulting operator/defect is topological. Although such lower-form symmetries act trivially via linking, they act \emph{faithfully} on higher-dimensional operators, as discussed in Section \ref{necessary conditions}.

\subsection{Gauge theory examples}\label{gaugetheoryexamples}

As we will show below, symmetry transmutation can happen in different ways. In particular, it is typical in gauge theories where the matter field content satisfies a gauge-flavor relation, as we explain in the following.

Consider first a UV gauge theory with gauge group $G_\mathrm{gauge}$.  If $G_\mathrm{gauge}$ does not act faithfully on the matter fields (e.g., there are no charged matter fields at all), the system has an electric one-form symmetry ${\cal Z}^{(1)}$ where ${\cal Z}$ is the subgroup of the center of $G_\mathrm{gauge}$ that does not act on the matter fields \cite{Gaiotto:2014kfa}.  Then, we can couple the UV gauge theory to background gauge fields for ${\cal Z}^{(1)}$.\footnote{Recall our notation: the group ${\cal Z}$  leads to a one-form symmetry ${\cal Z}^{(1)}$.} This has the effect of twisting the dynamical $G_\mathrm{gauge}$ gauge fields to be gauge fields of $G_\mathrm{gauge}\over {\cal Z}$ that are not gauge fields of $G_\mathrm{gauge}$.

We are interested in a situation without a UV one-form symmetry.  Therefore, we consider systems where the matter fields transform faithfully under $G_\mathrm{gauge}$.  However, it is commonly the case that we can still find such twisted $G_\mathrm{gauge}$ gauge fields as follows.  Let the matter fields transform under\footnote{Below, we will discuss more complicated situations where the product in the numerator can be a semi-direct product.  Also, in some cases, there are magnetic symmetries, which make the group theory richer.\label{can be more complicated}}
\ie
	\frac{G_\mathrm{naive} \times G_\mathrm{gauge}}{\cal Z}\,. \label{extended.group}
\fe
Here $G_\mathrm{naive}$ is the naive global symmetry group.  The quotient by a subgroup ${\cal Z}\subset G_\mathrm{gauge} $ of the center of the gauge group implies that not all representations of $G_\mathrm{naive} \times G_\mathrm{gauge}$ can be constructed out of the matter fields.  This means that the gauge invariant operators made out of the fundamental fields are in representations of
\ie\label{globalsymmetrya}
	\frac{G_\mathrm{naive} }{\cal Z}\,.
\fe

Let us demonstrate this abstract discussion in a well-known example.  In QCD with a single massive quark, the gauge group is $G_\mathrm{gauge}=\SU(N)$, and the naive global symmetry is the quark number symmetry $G_\mathrm{naive}=\U(1)_q$ under which the fundamental quarks carry charge one.  In this case, ${\cal Z}=\bZ_N$, the group in \eqref{extended.group} is $\U(N)={\U(1)_q\times \SU(N)\over \bZ_N}$, and the global symmetry that acts on the gauge invariant operators \eqref{globalsymmetrya} is the baryon number symmetry $\U(1)_B={\U(1)_q\over \bZ_N}$ under which the baryons have charge one.  Now, we can couple the system to background gauge fields for $\U(1)_B$ that are not background gauge fields of $\U(1)_q$.  As is clear from \eqref{extended.group}, such background gauge fields lead to twisted $\SU(N)$ bundles.  Even though the UV theory does not have a one-form global symmetry, we can still force it to have twisted $\SU(N)$ bundles.  Then, in phases where the UV zero-form symmetry $\U(1)_B$ does not act in the IR, it could happen that the IR theory exhibits a $\bZ^{(1)}_N$ one-form symmetry.  We say that the UV zero-form $\U(1)_B$ was transmuted in the IR into $\bZ_N^{(1)}$.  See Section \ref{sec:QCD}, for more details.

More generally, nontrivial background fields for the zero-form symmetry \eqref{globalsymmetrya} lead to nontrivial gauge bundles of $G_\mathrm{gauge}$.  In other words, such background fields for a zero-form symmetry act in the UV like background fields for an electric one-form symmetry ${\cal Z}^{(1)}$, even though the UV theory does not have such a one-form global symmetry.  Then, in phases where some of the UV zero-form symmetry does not act in the IR, it can be transmuted into the one-form symmetry ${\cal Z}^{(1)}$.

In the following sections, we will see many examples of such phenomena.  In particular, in Section \ref{O3d3}, we will study QED$_3$ with two charge-one scalars.  This UV system does not have any higher-form symmetry.  It exhibits many interesting IR phases.  In several of them, the IR theory has new one-form symmetries that resulted from symmetry transmutation, along the lines mentioned above.  In Table \ref{tab:summary}, we summarize some of these results.  We refer the reader to Section \ref{O3d3} for a more complete description of the notation and the various phenomena we discuss.

\begin{table}[t]\centering
\begin{tabular}{|c|c|c|c|}
\hline
Sec. & IR Phase &$G_\IR $ & $K^{(1)}$ \\ \hline
\ref{subsec:higgs} & $\mathbb{CP}^1$ (Higgs)& $\SO(3)_\mathrm{f} \times \O(2)$ & - \\ \hline
\ref{sec:coloumb} & $S^1$ (Coulomb)  & $\left( \U(1)_\mathrm{e}^{(1)} \times \U(1)_\mathrm{m} \right) \rtimes \bZ_2^C$ & $\bZ_{2,\mathrm{e}}^{(1)}$ \\ \hline
\ref{sec:so3.sigma} & SO(3) sigma model &  $\SO(3)_\mathrm{f} \times \bZ_{2}^{(1)} \times \bZ_2^C$ & $\bZ_{2}^{(1)}$ \\ \hline
\ref{subsec:z2} & $\bZ_2$ gauge theory & $\bZ_{2,\mathrm{e}}^{(1)} \times \bZ_{2,\mathrm{m}}^{(1)}$ & $\bZ_{2,\mathrm{e}}^{(1)} \times \bZ_{2,\mathrm{m}}^{(1)}$ \\ \hline
\ref{subsec:u1k.cs} & U(1)$_k$ Chern-Simons & $\bZ_k^{(1)} \rtimes \bZ_2^C$ & $\bZ_k^{(1)} $ \\ \hline
\end{tabular}
\caption{Various IR phases of two-flavor scalar QED$_3$.  The UV zero-form symmetry is $G_\UV  = \SO(3)_\mathrm{f} \times \O(2)$. Different IR phases have different patterns of symmetry breaking and symmetry transmutation. The faithful symmetry of the infrared theory is denoted as $G_\IR $.  It includes both zero-form and one-form symmetries. Finally, $K^{(1)} \subset G_\IR $ in the last column, is the IR one-form symmetry that resulted from transmutation.  The rest of the IR one-form symmetry is emergent.}
\label{tab:summary}
\end{table}

\subsection{Outline}

In Section \ref{sec:2d.u1}, we examine two phases of two-flavor scalar QED in 1+1 dimensions. In particular, Section \ref{sec:qed3.massive} focuses on a massive phase with symmetry transmutation from a zero-form O(3) symmetry into a $\bZ_2^{(1)}$ one-form symmetry. Section \ref{sec:lattice} presents an exactly solvable lattice model in 1+1d with symmetry transmutation from $\bZ_2 \times \bZ_2$ zero-form symmetry into a $\bZ_2^{(1)}$ one-form symmetry. In Section \ref{O3d3}, we discuss various phases of two-flavor scalar QED in 2+1d and their corresponding symmetry transmutation patterns. Section \ref{sec:QCD} explores symmetry transmutation in one-flavor QCD in 3+1 dimensions and its connection to confinement.

Section \ref{sec:general} offers various general remarks on symmetry transmutation. In particular, Section \ref{sec:central extension} elaborates on the symmetry transmutation map and its relation to the projective action of the zero-form symmetry on the UV line operators. Section \ref{A simple example} presents a simple model that demonstrates it. Section  \ref{sec:extension.emergent.anomalies} discusses an application of symmetry transmutation and discusses emergent anomalies.

We conclude our findings in Section \ref{Conclusions}, where we contrast symmetry transmutation with similar phenomena and discuss further generalizations.

In Appendix \ref{app:qm}, we discuss anomalies of 1-form symmetries in quantum mechanics. These anomalies will be important in our analysis of higher-dimensional field theories.  Appendix \ref{app:math} provides further details on the mathematical nature of the symmetry transmutation map.

\section{1+1d U(1) gauge theory} \label{sec:2d.u1}

As the first example, we study two-flavor scalar QED in two spacetime dimensions with the Lagrangian (in Euclidean signature)
\ie\label{QED2UVL}
	\mathcal{L}_\mathrm{Euclidean} = \frac{1}{4e^2} f_{\mu\nu}f^{\mu\nu} - i\frac{\theta}{4\pi} \varepsilon_{\mu\nu}f^{\mu\nu} +  | D_a \phi^i |^2 + m^2 |\phi^i|^2 + V(|\phi^i|^2) \,,
\fe
where $f_{\mu\nu} = \partial_\mu a_\nu - \partial_\nu a_\mu$ is the U(1)$_\mathrm{EM}$ field strength, and $\phi^1$ and $\phi^2$ are complex scalar fields. The U(1)$_\mathrm{EM}$ gauge field $a$ is normalized such that $\int \mathrm{d}a = \int \frac{1}{2} f_{\mu\nu} \varepsilon^{\mu\nu} \mathrm{d}^2x  \in 2\pi \bZ$, therefore $\theta$ is $2\pi$ periodic. We focus mostly on the theory with $\theta=\pi$.

\subsection{The UV theory: the global symmetry and its anomaly}\label{U1UV}

In terms of the discussion in Section \ref{gaugetheoryexamples}, for generic $\theta$, the gauge group is $G_\mathrm{gauge}=\U(1)_\mathrm{EM}$, the naive global symmetry is $G_\mathrm{naive}=\SU(2)$, and ${\cal Z}=\bZ_2$.  Consequently, as in \eqref{globalsymmetrya},
 the UV global symmetry is
\ie
	G_\UV  = \mathrm{SO}(3) \,.
\fe

For $\theta$ a multiple of $\pi$, $G_\mathrm{naive}$ includes also a $\bZ_2^C$ charge conjugation symmetry generated by the charge conjugation transformation $U_C : a \mapsto -a$ and $U_C: \phi^i \mapsto \varepsilon^{ij} \phi_j^{*}$ (where $\phi_j^*=(\phi^j)^*$). Note that $U_C^2 = -1 \in \U(1)_\mathrm{EM}$ and acts as $\phi^i \mapsto -\phi^i$.
Then, $G_\mathrm{naive}$ acts on the gauge group, and therefore the product in \eqref{extended.group} is actually a semi-direct product.  The result is that the global symmetry \eqref{globalsymmetrya} is
\ie\label{SO3X2C}
	G_\UV  = \mathrm{O}(3) = \mathrm{SO}(3)\times\bZ_2^C \,.
\fe

As in the discussion in Section \ref{gaugetheoryexamples}, the matter fields transform under the quotient
\ie\label{pinthreenaive}
	\frac{\mathrm{Pin}^-(3) \ltimes \U(1)_\mathrm{EM}}{\bZ_2}\,,
\fe
where $G_\mathrm{naive}=\mathrm{Pin}^-(3)$ is the group generated by SU(2) transformations and charge conjugation transformation $U_C$ such that $U_C^2$ is identified with the $2\pi$-rotation of SU(2), and hence acts as $\bZ_4^C\subset\mathrm{Pin}^-(3)$.\footnote{Note that this $\mathrm{Pin}^-(3)$ is an internal symmetry rather than a spacetime symmetry, which depends on the signature.} In other words, the matter fields transform projectively under O(3), but linearly under $\mathrm{Pin}^-(3)$. The semi-direct product above captures the action of the charge conjugation symmetry $\bZ_4^C$ on $\U(1)_\mathrm{EM}$. Finally, the $\bZ_2$ in the denominator of the quotient above corresponds to $2\pi$-rotations of SU(2) and $U_C^2 = -1 \in \U(1)_\mathrm{EM}$. As a consequence, a twisted $\O(3) = \frac{\mathrm{Pin}^-(3)}{\bZ_2}$ bundle with nonzero $w_2[\SO(3)]$ and background gauge field $C$ for $\bZ_2^C$ leads to fractional flux of $a$:\footnote{Here $w_2[\mathrm{SO}(3)] \in H^2(\Sigma, \bZ_2)$ is a characteristic class of the SO(3) bundle that obstructs lifting it to an SU(2) bundle. \label{w2def}}
\ie
	\left (2\int {\mathrm{d}a\over 2\pi} \right) \mod 2 = \int \big( w_2[\SO(3)] + C \cup C \big) \,. \label{bundle.relation}
\fe
This means that even though our system does not have a one-form symmetry, nontrivial $\O(3)$ background gauge fields act effectively as $\bZ_2^{(1)}$ one-form symmetry backgrounds.\footnote{The relation \eqref{bundle.relation}, is similar to the spin/charge relations in spin$^c$ theories.  For its application in physical systems, see e.g., \cite{Metlitski:2015yqa, Seiberg:2016rsg}.}

Let us discuss it in more detail.  The charge conjugation transformation $U_C$ changes $\theta$ to $-\theta$. Thus, at $\theta=\pi$, it needs to be accompanied by the unitary transformation $e^{i \int_\text{space} a}$ that shifts $\theta$ by $2\pi$ and brings $\theta=-\pi$ back to $\theta=\pi$. Thus, $U_C(\theta = \pi) = U_C(\theta = 0) e^{i \int_\text{space} a}$. This extra unitary transformation introduces an 't Hooft anomaly in the O(3) symmetry\footnote{This theory and its anomaly have been discussed in \cite{Gaiotto:2017yup, Komargodski:2017dmc}.}
\ie
	\alpha_\UV  = \pi i \int (w_2[\mathrm{SO}(3)] + C^2) \cup C \,. \label{qed2.anomaly}
\fe
The $C^2 \cup C$ term represents the nontrivial element of $H^3(\bZ_2 ,U(1)) \simeq \bZ_2$ that describes the pure $\bZ_2$ anomaly in 1+1d.\footnote{Our notation is $C^2 = C\cup C = \frac{\delta \tilde{C}}{2}$, where $\tilde{C} \in C^1(\Sigma,\bZ_4)$ is a $\bZ_4$ lift of the $\bZ_2$ gauge field $C  \in Z^1(\Sigma,\bZ_2)$. Here, $C^1(\Sigma,\bZ_4)$ denotes $\bZ_4$ cochains, which are not necessarily closed, and $Z^1(\Sigma,\bZ_2)$ denotes $\bZ_2$ closed cochains, i.e., cocycles. Similarly we have $C^3 = C\cup C \cup C = \frac{\delta \tilde{C}}{2} \cup C$.}

To better understand the anomaly, we place the theory on a circle with $\bZ_2^C$-twisted boundary condition, i.e., insert a $\bZ_2^C$ defect on a point on the circle. (The original $\bZ_4^C \subset\mathrm{Pin}^-(3)$ symmetry became $\bZ_2^C$ after the $\bZ_2$ quotient in \eqref{pinthreenaive}.)  The anomaly theory \eqref{qed2.anomaly} implies that the effective QM theory on a $\bZ_2^C$-twisted circle is in a projective representation of O(3) described by $\pi \int w_2[\mathrm{SO}(3)] + C^2$. The first term corresponds to half-integer representations of SO(3), and the second term corresponds to $(U_C)^2 = -1$. Below, we verify the anomaly by reproducing these projective representations.

As we said above, at $\theta=\pi$, the operator/defect $U_C$ should include a factor of $e^{i\int a}$.  Therefore, the $\bZ_2^C$ defect includes a time-like Wilson line $e^{i \int_\text{time} a}$.   This factor describes the worldline of a particle with U(1)$_\mathrm{EM}$ charge one. Consequently, all the states in the Hilbert space of the $\bZ_2^C$-twisted theory have charge-one under U(1)$_\mathrm{EM}$.  Since the fundamental charge-one fields, $\phi_1$ and $\phi_2$, are doublets of $\mathrm{Pin}^-(3)$, the Hilbert space transforms projectively under the $O(3)$ global symmetry, thus reflecting the anomaly theory \eqref{qed2.anomaly}.

\subsection{The massive phase -- O(3)$^{(0)}$ transmuted into $\bZ_2^{(1)}$\label{sec:qed3.massive}}

Here, we study the phase of the theory at $\theta=\pi$ for $m^2 \gg e^2$.

\subsubsection{The IR theory}

We now study the IR theory at energies below the mass of the scalar fields $m$, and relate the global symmetry of the UV theory to the symmetry of the IR theory. After integrating out the massive scalars, we find the pure U(1)$_\mathrm{EM}$ gauge theory in one spatial dimension, with the action (in Lorentzian signature)
\ie
	S_\mathrm{Lorentzian} = \int  \left( \frac{1}{2e^2} (f_{01})^2 + \frac{\theta}{2\pi} f_{01} \right) \mathrm{d}^2x \,. \label{QED2IRL}
\fe	
The equation of motion sets the electric field $f_{01}$ to a constant.  Since it is a \emph{local} conserved quantity (i.e.,\ a topological local operator) it generates a U(1)$^{(1)}$ \emph{one-form} symmetry \cite{Gaiotto:2014kfa}.  The corresponding integer-valued charge is
\ie
	Q = \frac{1}{e^2} f_{01} + \frac{\theta}{2\pi} \in \bZ \,.
\fe	
The operators charged under this one-form symmetry are the Wilson lines $e^{i\int a}$.

For $\theta \in \pi\bZ$, there is also a $\bZ_2^C$ charge conjugation zero-form symmetry
acting as
\ie
	U_C \, Q \, U_C^{-1} = \frac{\theta}{\pi} - Q \qquad,\qquad \theta\in \pi\bZ \,.
\fe
For $\theta\in \pi(2\bZ+1)$, this leads to a projective representation of the algebra of $\bZ_2^C$ and the U(1)$^{(1)}$ one-form symmetry.  Specifically, for $\theta=\pi$
\ie
e^{i \alpha Q}  U_C=e^{i\alpha} U_C e^{-i \alpha Q}
\,. \label{projective.alg}
\fe
(Alternatively, we can study ${\cal U}_\alpha=e^{i\alpha(Q-{1\over 2})}$ such that the only projective phase
is in ${\cal U}_{2\pi}=-1$.)

This projective phase signals a mixed 't Hooft anomaly between U(1)$^{(1)}$ and $\bZ_2^C$, and it implies a degeneracy in the spectrum of the theory \cite{Gaiotto:2017yup}. In particular, focusing on the sub-symmetry of the one-form symmetry $\bZ_2^{(1)} \subset \U(1)^{(1)}$ generated by $e^{i\pi Q}$, and the zero-form symmetry $\bZ_2^C$, the 2+1d anomaly theory is given by\footnote{As we said in footnote \ref{BvscalB}, we mostly use $B$ for a U(1) two-form gauge field and ${\cal B}$ for $\bZ_N$ two-form gauge fields, and when they are related, $\int B={2\pi \over N} \int {\cal B}$.  The periods of $B$ are real modulo $2\pi$ and the periods of $\cal B$ are integers modulo $N$.}
\ie
	\alpha_\IR  = \pi i \int {\cal B}_\IR  \cup C \,, \label{em.anomaly}
\fe	
where ${\cal B}_\IR $ is a discrete background gauge field for the $\bZ_2$ subgroup of the $\U(1)^{(1)}$ 1-form symmetry and $C$ is a background for the $\bZ_2^C$ charge conjugation symmetry.
For $\theta = 0$, there is no projective phase in \eqref{projective.alg}, and therefore, there is no anomaly.

Putting the theory on a circle $S^1$ of radius $R$ leads to an effective QM system with the Hamiltonian
\ie\label{energiesopo}
	H = \frac{ 2\pi R e^2}{2} \left( Q - \frac{\theta}{2\pi} \right)^2 \,.
\fe

The spectrum \eqref{energiesopo} points to an important subtlety that we should stress.  When we study the low-energy theory, we have in mind placing the system in large, but finite volume.  In our case, this means that we should take $R$ to be large.  Our system is gapped, and for $Re^2 \gg m$, the excited states of \eqref{QED2IRL} lie beyond the regime of validity of this effective low-energy theory. Their energies are even higher than the localized states of the UV theory \eqref{QED2UVL}.  Therefore, the spectrum \eqref{energiesopo} cannot be interpreted as the spectrum of the low-energy theory.  Only the ground states in \eqref{energiesopo} are meaningful.\footnote{This is not to say that the Lagrangian \eqref{QED2IRL} is meaningless, as it can be studied as a complete theory and then the spectrum \eqref{energiesopo} is valid.  Also, even as a low-energy effective theory of some higher-energy theory, for some range of parameters, it can capture properties of confining strings that can break in the full theory.}

For generic $\theta$, there is no $\bZ_2^C$ symmetry and the ground state is unique.  In this case, the true low-energy theory is trivial.  For $\theta=0\mod 2\pi$, the theory has a $\bZ_2^C$ symmetry and it is unbroken in a unique ground state.  Again, the low-energy theory is trivial.  For $\theta=\pi\mod 2\pi$, the theory has a $\bZ_2^C$ symmetry and it is spontaneously broken.  In this case, there are two low-lying states and the low-energy theory includes only these two states.

Let us discuss the effective theory of these two states $|\pm\rangle $.  They are labeled by the eigenvalue of the order parameter $\cal O$ for the spontaneously broken $\bZ_2^C$ symmetry, ${\cal O}|\pm\rangle=\pm|\pm\rangle$.  Since for $\theta=\pi$, these states have $Q=0, 1$, we can identify ${\cal O}=e^{i\pi Q}$.  This identification is consistent with the symmetry algebra \eqref{projective.alg}.  We see that the order parameter of the spontaneously broken $\bZ_2^C$ symmetry, ${\cal O}=e^{i\pi Q}$, generates a $\bZ_2^{(1)}\subset \U(1)^{(1)}$ one-form symmetry. (Importantly, the entire $\U(1)^{(1)}$ one-form symmetry acts in the low-energy theory and not only its $\bZ_2^{(1)}$ subgroup.)

The same picture of the Hilbert space arises in a $\bZ_2$ gauge theory with Wilson line $U_C$.  Its two-dimensional Hilbert space includes the same two states as above, and its $\bZ_2^{(1)}$ one-form symmetry is the one generated by ${\cal O}$.  However, in this case, the $\bZ_2$ gauge theory variables allow us to express only the $\bZ_2^{(1)}$ one-form symmetry. Operators of the form $e^{i\alpha {\cal O}}$ realize the full $\U(1)^{(1)}$, but they are not natural in the $\bZ_2$ gauge theory.

Let us continue to study the full U(1) gauge theory \eqref{QED2IRL} and not only its low-energy limit and add to it a $C$ defect.   This means that we have background $C$ such that $(-1)^{\int_{S^1} C} = -1$. Consequently, for any local operator, $O(x+2\pi R) = U_C \, O(x) \, U_C^{-1}$. Since $Q$ is a constant, the twisted boundary condition leads to the identification
\ie
	Q = \frac{\theta}{\pi} - Q \,, \label{id}
\fe
and hence, $Q = \frac{\theta}{2\pi}$. Moreover, since $Q$ is quantized, this is possible only for $\theta\in 2\pi \bZ$.  In other words, for $\theta = 0\mod 2\pi$, there is only one state with $Q=0$ and for $\theta=\pi$, the QM theory has no states!

The lack of states in the Hilbert space can also be explained by reducing the anomaly \eqref{em.anomaly} on the circle. The nontrivial background of $C$ on the spatial circle leads to the anomaly $(-1)^{\int {\cal B}_\IR }$ for the effective QM theory at $\theta=\pi$. As discussed in Appendix \ref{app:qm}, this anomaly implies that the Hilbert space of the QM theory has no state.

\begin{figure}[h]
\centering
\includegraphics[width=10cm]{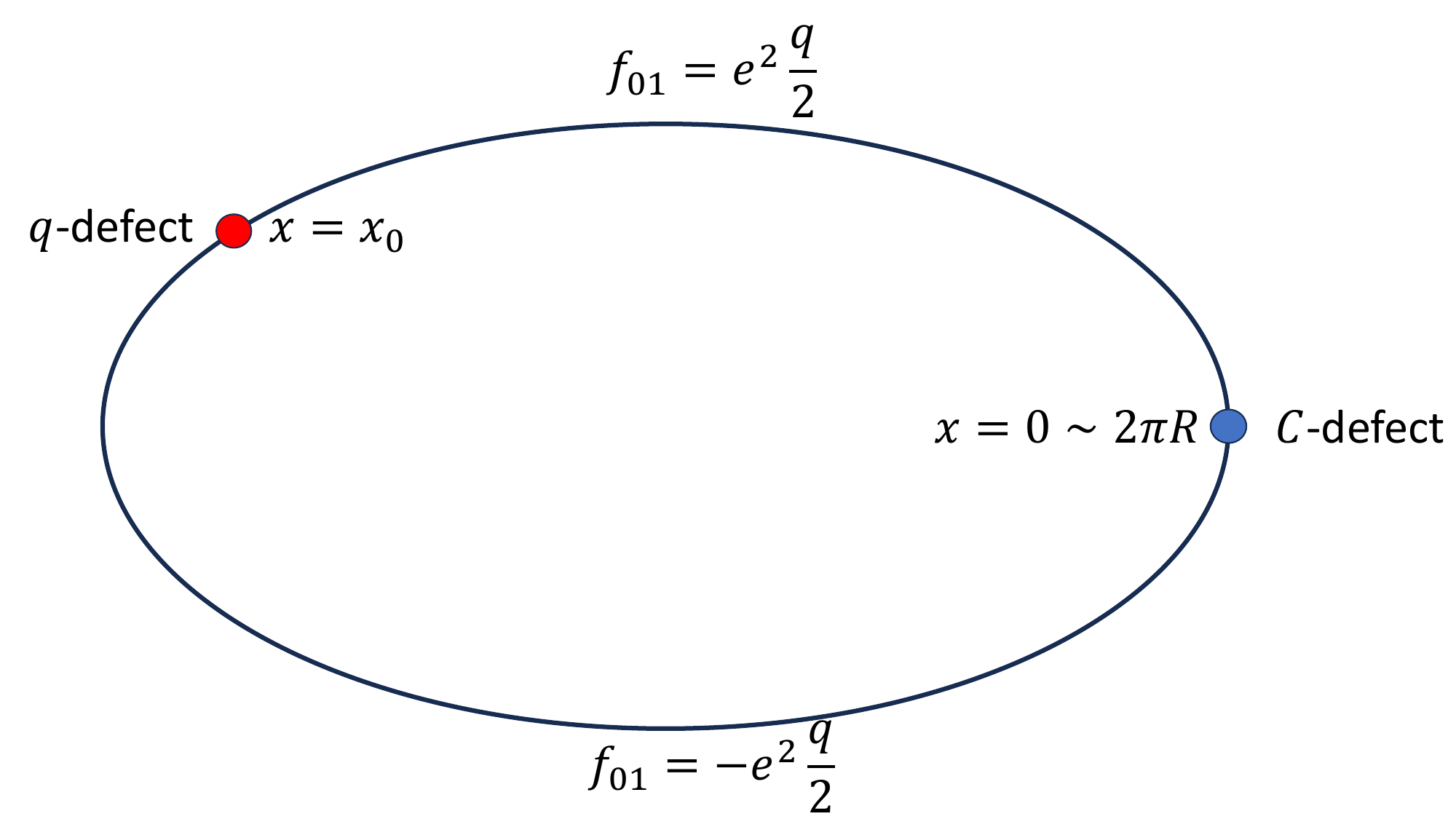}
\caption{The configuration of the spatial circle with a (topological) charge conjugation defect at $x=0$ (the blue point) and a non-topological defect of charge $q$ at $x=x_0$ (the red point). The latter is given by $W$ of \eqref{Wqdef}.  The electric field between the two defects is $f_{01} = e^2(Q-\frac{\theta}{2\pi}) = \pm e^2 \frac{q}{2}$. Note that for $\theta=0$ and even $q$, the Hilbert space has a single state.  And for odd $q$, the Hilbert space is empty. For $\theta=\pi$,  there are no states unless we add a defect with an odd charge $q$.}
\label{circle.with.defect}
\end{figure}

To get a nonempty Hilbert space for a theory with a $C$ defect, we can add a non-topological defect that is also charged under the one-form symmetry. From the point of view of the IR theory, this is a probe particle of odd charge $q \in 2\bZ+1$.   Let it be at point $x_0$.  It can be represented by a time-like Wilson line
\ie\label{Wqdef}
W=e^{i q \int_{x=x_0} a} ~ .
\fe
See Figure \ref{circle.with.defect}. Importantly, the $C$ defect breaks the $\U(1)^{(1)}$ one-form symmetry to $\bZ_2^{(1)}$.  Therefore, the action  of the unbroken one-form symmetry depends only on $q\mod 2$. Indeed, the UV operator $e^{i\int_0^{x_0}a -i\int_{x_0}^{2\pi R} a}$ has the effect of shifting $q$ of the defect by $2$.  Consequently, in the IR theory, all the defects with the same $q\mod 2$ should be identified.

This defect corresponds to adding $\delta \mathcal{L} = q \delta(x-x_0) a_t$ to the Lagrangian, which modifies Gauss's law to $\partial_x Q = q \delta(x-x_0)$. In other words, the defect has charge $q$ under the one-form symmetry and therefore
\ie
	Q(x > x_0) = Q(x < x_0) + q \,.
\fe	
This relation and the identification \eqref{id} from the presence of the $C$ defect leads to $Q = \frac{\theta}{2\pi} \pm q/2$.
From the quantization of $Q$, we find that the effective QM theory has a unique state of energy $E = \pi R q^2 e^2 /4$ if $\theta = \pi q + 2 \pi \bZ$.

The fact that the Hilbert space of the theory at $\theta=\pi$ must carry charge one under the gauge group is consistent with our discussion in the full theory at the end of Subsection \ref{U1UV}.

\subsubsection{The relation between the UV and IR and symmetry transmutation}

Specializing to $\theta=\pi$, we compare the UV and the IR theories. They have different faithful global symmetries
\ie
	G_\UV  = \mathrm{SO}(3)\times \bZ_2^C \qquad \text{and} \qquad G_\IR  = \mathrm{U}(1)^{(1)} \rtimes \bZ_2^C \,.
\fe
Here, we will explain the relation between $G_\UV $ and $G_\IR $ and match their 't Hooft anomalies.

At energies below the mass of the scalars, the SO(3) global symmetry does not act on gauge invariant local operators. However, as we will discuss, it does act nontrivially on line operators and effectively becomes a one-form symmetry.  It is transmuted into the one-form symmetry $\bZ_2^{(1)}\subset \U(1)^{(1)}$ generated by ${\cal O}=e^{i\pi Q}$ with background gauge field
\ie
	{\cal B}_\IR  = w_2[\mathrm{SO}(3)] + C^2 \,. \label{uv.ir.map}
\fe
The UV anomaly \eqref{qed2.anomaly} is matched by substituting \eqref{uv.ir.map} into the IR anomaly \eqref{em.anomaly}. Equation \eqref{uv.ir.map} describes the relation between the UV and IR gauge fields and replaces the homomorphism \eqref{homomorphism} of zero-form symmetries or the corresponding homomorphism between gauge fields of zero-form symmetries \eqref{ordinary.map.Phi}.

Note that, unlike the SO(3) symmetry, the charge conjugation $\bZ_2^C$ acts faithfully in the IR. Consequently, we can redefine ${\cal B}_\IR $ to ${\cal B}_\IR  - C^2$. This redefinition changes the map \eqref{uv.ir.map} and also shifts the IR anomaly \eqref{em.anomaly} by the pure 1+1d $\bZ_2$ anomaly $C^3$ as discussed in \cite{Delmastro:2022pfo}.

In the following, we will discuss two different presentations of symmetry transmutation and the relation \eqref{uv.ir.map}. They all rely on the fact that the O(3) symmetry acts projectively on the UV matter fields.

\begin{enumerate}
\item \textbf{Gauge/flavor relation:} We note that the matter fields transform under
\ie
	\frac{\mathrm{Pin}^-(3) \ltimes \U(1)_\mathrm{EM}}{\bZ_2} = \bZ_2^C \ltimes \mathrm{U}(2)
\fe
(compare with \eqref{extended.group}), and the faithful global symmetry is given by the quotient
\ie
	G_\UV  = \frac{\mathrm{Pin}^-(3)}{\bZ_2} = \mathrm{O}(3) \,. \label{1+1d.quotient}
\fe
As a result, the matter fields with U(1)$_\mathrm{EM}$ charge one transform in a projective representation of O(3), which is a linear representation of $\mathrm{Pin}^-(3)$. This leads to the relation \eqref{bundle.relation} between the characteristic classes of O(3) background gauge fields and the U(1)$_\mathrm{EM}$ gauge bundle.
We recognize the left-hand side of \eqref{bundle.relation} as the holonomy of the IR $\bZ_2^{(1)}\subset \U(1)^{(1)}$ symmetry.  This shows that we can turn on a background gauge field for the IR $\bZ_2^{(1)}\subset \U(1)^{(1)}$ symmetry by turning on a background gauge field for the UV O(3) symmetry.

\item \textbf{Geometric picture:} Another way to see the symmetry transmutation is to note that an intersection of UV symmetry defects becomes a higher-form symmetry operator/defect in the IR. This is mainly because the O(3) global symmetry of the UV acts projectively on the Wilson lines. More precisely, the defect Hilbert space of a charge-one Wilson line transforms in a projective representation of O(3) described by the right-hand side of \eqref{uv.ir.map}.  (Compare with the discussion at the end of Subsection \ref{U1UV}.)

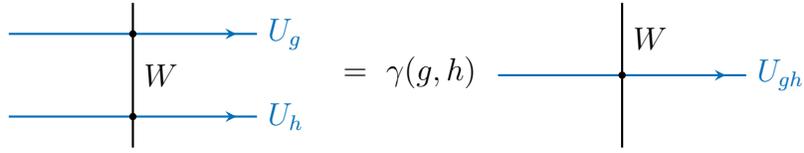
\begin{figure}[t]
    \centering
    \raisebox{-26pt}{\begin{tikzpicture}[scale=0.55]
	
        \draw[thick, NavyBlue] (-3,-1) -- (3,-1);
        \draw[thick, NavyBlue, -stealth] (2.4,-1) -- (2.5,-1);
        \draw[thick, black] (0,-1.75) -- (0,1.75) node[pos=0.5, right] { $W$};
        \draw[thick, NavyBlue] (-3,1) -- (3,1);
         \draw[thick, NavyBlue, -stealth] (2.4,1) -- (2.5,1);
	
        \fill[black] (0,1) circle (2.5 pt);
        \fill[black] (0,-1) circle (2.5 pt);

        \node[right] at (3,-1) { \color{NavyBlue} $U_h$};
        \node[right] at (3,1) { \color{NavyBlue} $U_g$};
    \end{tikzpicture}} $~=~ \gamma(g,h) ~$
    \raisebox{-26pt}{\begin{tikzpicture}[scale=0.55]
	
	\draw[thick, NavyBlue] (-3,0) -- (3,0);
	  \draw[thick, NavyBlue, -stealth] (2.4,0) -- (2.5,0);
        \draw[thick, black] (0,-1.75) -- (0,1.75) node[pos=0.75, right] { $W$};
	
        \fill[black] (0,0) circle (2.5 pt);

        \node[right] at (3,0) { \color{NavyBlue} $U_{gh}$};
    \end{tikzpicture}}
    \caption{Projective action of a zero-form symmetry on a line operator/defect $W$. Symmetry operators $U_g$, for $g \in G$, can act projectively on the $W$-defect Hilbert space. The projective action is characterized by $[\gamma] \in H^2(G,U(1))$, where $U_gU_h = \gamma(g,h) \, U_{gh}$. The phase $\gamma(g,h) \in U(1)$ is localized at the intersection of $W$ with symmetry operators.}
    \label{fig:2d.proj.action.on.lines}
\end{figure}

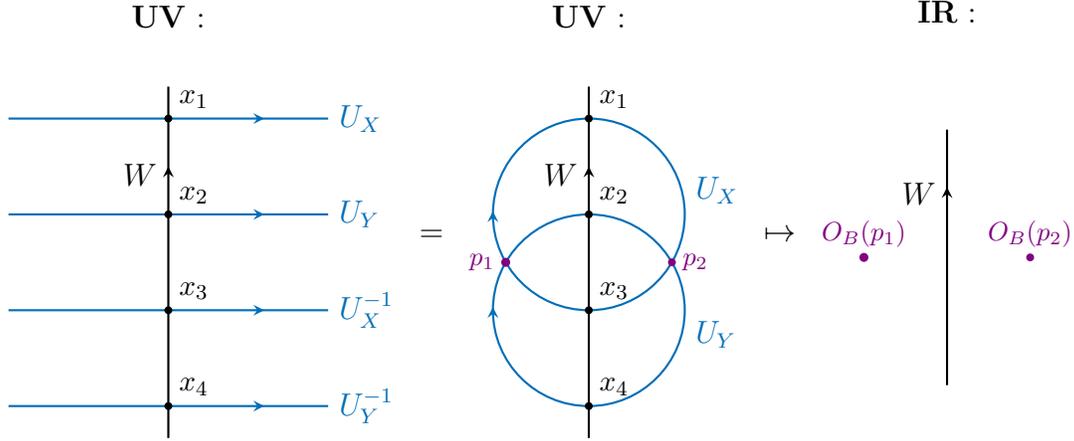
\begin{figure}[t]
    \centering
    \raisebox{-75pt}{\begin{tikzpicture}[scale=0.85]

     \node[above] at (0,3.5) {\textbf{UV} :};

    \draw[thick, NavyBlue] (-2.5,2.25) -- (2.5,2.25);
    \draw[thick, NavyBlue, -stealth] (1.4,2.25) -- (1.5,2.25);
    \node[right] at (2.5,2.25) { \color{NavyBlue} $U_X$};
    \draw[thick, NavyBlue] (-2.5,0.75) -- (2.5,0.75);
    \draw[thick, NavyBlue, -stealth] (1.4,0.75) -- (1.5,0.75);
    \node[right] at (2.5,0.75) { \color{NavyBlue} $U_Y$};
    \draw[thick, NavyBlue] (-2.5,-2.25) -- (2.5,-2.25);
    \draw[thick, NavyBlue, -stealth] (1.4,-2.25) -- (1.5,-2.25);
    \node[right] at (2.5,-2.25) { \color{NavyBlue} $U_Y^{-1}$};
    \draw[thick, NavyBlue] (-2.5,-0.75) -- (2.5,-0.75);
    \draw[thick, NavyBlue, -stealth] (1.4,-0.75) -- (1.5,-0.75);
    \node[right] at (2.5,-0.75) { \color{NavyBlue} $U_X^{-1}$};

    \draw[thick, black] (0,-2.75) -- (0,2.75) node[pos=0.75, left] { $W$};
    \draw[thick, black,-stealth] (0,1.4) -- (0,1.5);

    \fill[black] (0,2.25) circle (1.75pt); \node[above right] at (0,2.25) {\small $x_1$};
    \fill[black] (0,0.75) circle (1.75pt); \node[above right] at (0,0.75) {\small  $x_2$};
    \fill[black] (0,-2.25) circle (1.75pt); \node[above right] at (0,-0.75) {\small  $x_3$};
    \fill[black] (0,-0.75) circle (1.75pt); \node[above right] at (0,-2.25) {\small  $x_4$};
\end{tikzpicture}} $\,=\,$
	\raisebox{-75pt}{\begin{tikzpicture}[scale=0.85]
	\node[above] at (0,3.5) {\textbf{UV} :};

    \draw[thick, color=NavyBlue] (0,-0.75) circle (1.5);
     \draw[thick, NavyBlue, -stealth] (-1.5,0.7) -- (-1.5,0.8);

    \draw[thick, color=NavyBlue] (0,0.75) circle (1.5);
    \draw[thick, NavyBlue, -stealth] (-1.5,-0.8) -- (-1.5,-0.7);
    \node[above right] at (1.5,0.75) { \color{NavyBlue} $U_{X}$};
    \node[below right] at (1.5,-0.75) { \color{NavyBlue} $U_{Y}$};

    \draw[thick, black] (0,-2.75) -- (0,2.75) node[pos=0.75, left] { $W$};
    \draw[thick, black,-stealth] (0,1.4) -- (0,1.5);

    \fill[black] (0,2.25) circle (1.75pt); \node[above right] at (0,2.25) {\small $x_1$};
    \fill[black] (0,0.75) circle (1.75pt); \node[above right] at (0,0.75) {\small  $x_2$};
    \fill[black] (0,-2.25) circle (1.75pt); \node[above right] at (0,-0.75) {\small  $x_3$};
    \fill[black] (0,-0.75) circle (1.75pt); \node[above right] at (0,-2.25) {\small  $x_4$};
    \fill[violet] (1.3,0) circle (1.75 pt);  \node[right] at (1.3,0) {\footnotesize \color{violet} $p_2$};
    \fill[violet] (-1.3,0) circle (2pt); \node[left] at (-1.3,0) {\footnotesize \color{violet} $p_1$};
\end{tikzpicture}} $\,\mapsto\,$ \raisebox{-55pt}{\begin{tikzpicture}[scale=0.85]
	
    \node[above] at (0,3.5) {\textbf{IR} :};
	
    \draw[thick, black] (0,-2) -- (0,2) node[pos=0.75, left] { $W$};
    \draw[thick, black,-stealth] (0,1) -- (0,1.1);

    \fill[violet] (1.3,0) circle (1.75 pt);  \node[above] at (1.3,0) {\footnotesize \color{violet} $O_B(p_2)$};
    \fill[violet] (-1.3,0) circle (2pt); \node[above] at (-1.3,0) {\footnotesize \color{violet} $O_B(p_1)$};
\end{tikzpicture}}
    \caption{Transmuting a zero-form symmetry to a one-form symmetry in terms of defects. We study the defect configuration in the middle figure.  It represents $\bZ_2^{X} \times \bZ_2^{Y}$ 0-form symmetry defects $U_X$ and $U_Y$ intersecting the Wilson line $W$ at the  points $x_1,x_2,x_3,x_4$, and intersecting each other at the points $p_1$ and $p_2$. As in the left figure, shrinking the symmetry lines evaluates to $U_XU_YU_X^{-1}U_Y^{-1}=-1$. The right configuration describes the IR picture, where the UV zero-form symmetry acts trivially, but it is transmuted into the $\bZ_2^{(1)}$ one-form symmetry operators $O_B(p_1)$ and $O_B(p_2)$ localized at the intersection points $p_1$ and $p_2$. Bringing the two one-form symmetry operators together must evaluates to $-1$, thus $O_B W O_B = - W$.}
    \label{fig:intersection.symm.lines}
\end{figure}

This corresponds to the configuration in Figure \ref{fig:2d.proj.action.on.lines}, for $W$ as in \eqref{Wqdef}. To establish \eqref{uv.ir.map}, we focus on the $\bZ_2^X \times \bZ_2^Y$ subgroup of SO(3) generated by $\pi$-rotations along the $x$ and $y$ axes, which acts projectively on the matter fields. If we turn on $\bZ_2$ background gauge fields, $\mathcal{A}_X$ and $\mathcal{A}_Y$, for these two $\bZ_2^{(0)}$ symmetries, the relation \eqref{uv.ir.map} becomes:
\ie
	{\cal B}_\IR  = \mathcal{A}_X \cup \mathcal{A}_Y \,. \label{uv.ir.map2}
\fe
This relation means that the intersection of $\bZ_2^{X}$ and $\bZ_2^{Y}$ symmetry defects is a $\bZ_2^{(1)}$ one-form symmetry operator of the infrared theory.

As we now explain, this fact follows from Figure \ref{fig:2d.proj.action.on.lines}.
Let us insert the $\bZ_2^{X} \times \bZ_2^{Y}$ symmetry defects along two circles that intersect at two points that link nontrivially with a Wilson line. (See the red points $p_1,p_2$ in Figure \ref{fig:intersection.symm.lines}.) To show that the intersections generate a one-form symmetry, we need to show that shrinking the circles gives the one-form symmetry charge of the Wilson line. We can easily see that by looking at the intersection of the symmetry defects with the Wilson lines at the four positions $x_1,x_2,x_3,x_4$ shown in Figure \ref{fig:intersection.symm.lines}. These four points correspond to the symmetry operator $U_X U_Y U_X^{-1} U_Y^{-1}$ that evaluates to $-1$ because of the projective action of the $\bZ_2^{X} \times \bZ_2^{Y}$ symmetry. We emphasize that a key assumption here is the fact that SO(3) symmetry does not act on the local operators of the IR theory.  Otherwise, the lines $U_X$ and $U_Y$ are nontrivial.

\end{enumerate}

Finally, let us compare the Hilbert space of the UV and the IR theory. In the presence of a $C$ defect, the UV theory at $\theta=0$ has the same zero-energy state as the IR theory. But for $\theta = \pi$, the Hilbert space has two ground states with electric fields $f_{01} = \pm \frac12 e^2$ and energy $E = \pi R e^2 /4$. More precisely,  there is a charge conjugation defect at some point on the circle and a dynamically generated U(1) charge of $q=\pm 1$ at another point.  The electric field between the defects is $ \pm \frac12 e^2$. The fact that the ground-state energy of the theory at $\theta = \pi$ is higher than that of $\theta=0$, explains why the corresponding Hilbert space is empty in the IR theory.

\subsubsection{Position-dependent $\theta$}\label{positiondep}

We end this subsection with a discussion of this theory with a space-dependent $\theta$.  Since $\theta$ is circle valued, we can have $\theta(x+2\pi R)= \theta(x)+2\pi N$ with integer $N$.  In that case, a careful definition of the theory shows that the total U(1) charge in space should be $N$ \cite{Cordova:2019jnf,Cordova:2019uob,Ferromagnets,Seiberg:2024yig}.  That charge is carried by $N$ quanta of $\phi^i$.  Therefore, if $N$ is odd, the Hilbert space of the system is in a projective representation of the global SO(3) symmetry. This fact can be phrased as an anomaly in the space of coupling constant \cite{Cordova:2019jnf,Cordova:2019uob} given by an expression similar to \eqref{qed2.anomaly}\footnote{$[X]_N$ denotes the reduction of a real form $X$ modulo $N$, such that $\int [X]_N = \left( \int X \right) \mod{N}$. \label{discrete.mod}}
\ie
	\alpha_\UV  =  i \pi \int_\mathcal{M} w_2[\mathrm{SO}(3)] \cup \left[ \frac{\mathrm{d} \theta}{2\pi} \right]_2 \,. \label{qed2t.anomaly}
\fe

We can repeat the discussion of the IR manifestation of the UV anomaly \eqref{qed2.anomaly} for the case of position-dependent $\theta$ with the anomaly \eqref{qed2t.anomaly}.  Since this problem does not have the charge-conjugation symmetry, we cannot have a charge-conjugation defect.  Instead, when $\theta$ winds around an odd number of times, the full Hilbert space of the theory is in a projective (half-integer spin) representation of SO(3).  Therefore, if we look for an IR theory with a ground state singlet, the Hilbert space is empty.  Then, as in Figure \ref{circle.with.defect}, we can add a non-topological defect, representing the nontrivial SO(3) representation to find a non-empty Hilbert space.

\subsection{$\mathbb{CP}^1$ phase \label{sec:CP1}}

Here, we discuss the theory at $\theta=\pi$ and with \emph{negative} $m^2$ and $|m^2| \gg e^2$. In this regime, the infrared theory is strongly coupled and is described by the $\SU(2)_1$ WZW model \cite{Affleck:1987ch}.

The faithful global symmetry of the IR theory is
\ie
	G_\IR  =\SO(4)=\frac{ \SU(2)_\mathrm{L} \times \SU(2)_\mathrm{R}}{\bZ_2}
\,, \label{g.ir.wzw}
\fe
with anomaly
\ie
	&\alpha_\IR [A_\mathrm{L},A_\mathrm{R}] = \\ &\frac{1}{4\pi} \int \Tr \left[ \left( A_\mathrm{L} \wedge \mathrm{d} A_\mathrm{L} -\frac{2i}{3} A_\mathrm{L}\wedge A_\mathrm{L} \wedge A_\mathrm{L}  \right) -  \left( A_\mathrm{R} \wedge \mathrm{d} A_\mathrm{R} -\frac{2i}{3} A_\mathrm{R}\wedge A_\mathrm{R} \wedge A_\mathrm{R} \right)  \right] . \label{anomaly.wzw}
\fe
where $A_\mathrm{L}$ and $A_\mathrm{R}$ are SO(4) background gauge fields.

There is no symmetry transmutation in this phase, and the anomaly is matched by an ordinary homomorphism of the zero-form symmetries from $G_\UV $ to $G_\IR $:
\ie
	\varphi: \SO(3) \times \bZ_2^C \to \SO(4)
\,.
\fe
Under this homomorphism, $\SO(3)$ is mapped into $\SO(3)\subset\SO(4)$ and $\bZ_2^C $ is mapped to the center of one of the two $\SU(2)$ subgroups of $\SO(4)$.
To see that $\varphi$ correctly reproduces the UV anomaly \eqref{qed2.anomaly}, we restrict the $\SO(3)$ symmetry in the UV to its $\bZ_2^X \times \bZ_2^Y$ subgroup with background gauge field $X$ and $Y$, which faithfully represents the UV anomaly as
\ie
	\alpha_\UV [X,Y,C] = \pi i \int \left( XY +X^2 + Y^2 + C\right) \cup C = \pi i \int (X+C)\cup(Y+C) \cup C\,.
\fe
This anomaly is usually referred to as the type-III anomaly \cite{deWildPropitius:1995cf,Wang:2014tia}. It has the property that $\bZ_2^C$ is the only anomalous $\bZ_2$ subgroup of $\bZ_2^X \times \bZ_2^Y \times \bZ_2^C$.  This is nicely matched by the IR anomaly, as the $\SU(2)_1$ WZW model has only a single anomalous $\bZ_2$ symmetry, namely the chiral $\bZ_2^\mathrm{L}$ symmetry.

\subsection{A simple related example based on a 1+1d CFT}\label{simpleexample}

A simple example demonstrating symmetry transmutation is the following.\footnote{We thank Shu-Heng Shao for suggesting this example.}  Consider the $c=1$ CFT in 1+1d.  It can be described by a compact scalar $\chi\sim \chi+2\pi$.  This model has a $\big( \U(1)_m\times \U(1)_w \big) \rtimes \bZ_2^{X}$ zero-form global symmetry.  We deform it by a potential term $\cos(2\chi)$, which breaks the zero-form symmetry to $G_\UV=\bZ_{2}^{m}\times \left(\U(1)_w\rtimes\bZ_2^{X}\right)$. Here, $\bZ_2^{m}$ and $\bZ_2^{X}$ act as $\chi \mapsto \chi+\pi $ and $\chi\mapsto -\chi$, respectively. As is well known, there is a mixed anomaly between $\bZ_2^{m}$ and $\U(1)^{}_w$ zero-form symmetries.

If the radius of the circle is larger than the self-dual radius, the cosine interaction is relevant, and the model is gapped in the IR with two vacua at $\langle \chi \rangle = \pm \pi/2$.  The unbroken symmetry in these two vacua is $\O(2)_w=\U(1)_w \rtimes \mathrm{diag}(\bZ_2^{X},\bZ_2^{m})\subset G_\UV$, where $\mathrm{diag}(\bZ_2^{X},\bZ_2^{m})$ acts as $\chi\mapsto -\chi+\pi$. Clearly, the $\bZ_{2}^{m}$ zero-form symmetry is spontaneously broken with the order parameter $\langle i e^{i\chi}\rangle=\pm 1$. This order parameter can be interpreted as the symmetry operator for an IR one-form symmetry $K_\IR^{(1)}=\bZ_2^{(1)}$. This IR symmetry is a result of symmetry transmutation from the UV symmetry $\O(2)_w$.  Clearly, the mixed anomaly between the UV symmetries $\bZ_{2}^{m}$ and $\U(1)_w$ is matched in the IR by a mixed anomaly between $\bZ_{2}^{m}$ and $\bZ_2^{(1)}$.  Indeed, the IR theory has defects associated with the broken $\bZ_{2}^{m}$ symmetry lines. They correspond to domain wall configurations, where $\chi$ winds halfway around its circle and hence, they have half-integer winding. More generally, the unbroken $\O(2)_w$ global symmetry  acts projectively on such configurations. As expected, the UV anomaly leads to symmetry fractionalization on these defects.

This example can also be viewed as a deformation of the two-flavor scalar QED$_2$ theory discussed earlier in this section. Starting in the UV, we can add to the $\SO(3)\times\bZ_2^C$ invariant Lagrangian \eqref{QED2UVL} the potential term
\ie\label{SO3bre}
|\phi^1\phi^2|^2\,,
\fe
which preserves $\bZ_2^C$ and explicitly breaks the global $\SO(3)$ symmetry group to $\O(2) = \U(1)_w \rtimes \bZ_{2}^{X} $, so we end up with the UV symmetry group $\bZ_2^C\times \left(\U(1)_w \rtimes \bZ_{2}^{X}\right)$. As we will soon see, this notation for the symmetry coincides with the notation above. $\U(1)_w$ acts as $\phi^1 \mapsto e^{i\alpha/2}\phi^1$, $\phi^2 \mapsto e^{-i\alpha/2}\phi^2$ and $\bZ_{2}^{X} $ exchanges $\phi^1\leftrightarrow \phi^2$. This symmetry group can also be rewritten as $\bZ_2^{m}\times \left(\U(1)_w \rtimes \bZ_{2}^{X}\right)$, where $\bZ_2^m = \mathrm{diag}(\bZ_2^C , \bZ_2^w)$.

With appropriate coefficients in the potential, the minima are at $\langle\phi^1\rangle=0$, $\langle\phi^2\rangle\ne0$ and at $\langle\phi^2\rangle=0$, $\langle\phi^1\rangle\ne0$. In these vacua, the $\U(1)_\mathrm{EM}$ gauge group is completely Higgsed and the global symmetry is spontaneously broken to $\U(1)_w \rtimes \mathrm{diag}(\bZ_2^X, \bZ_2^m)$, which is the same pattern of symmetry breaking as above.

\section{1+1d lattice model \label{sec:lattice}}

Consider a spin-$\frac12$ periodic chain with the Hamiltonian
\ie
	H = - \sum_{j =1}^L h_j \left( X_j + Z_{j-1} X_j Z_{j+1} \right) \,, \label{lattice.H}
\fe
where the total number of sites is an even integer $L$, and $X_j$ and $Z_j$ are the Pauli operators acting on the qubit at site $j$. The Hamiltonian has a
\ie\label{GUVlattice}
G_\UV=\bZ_2^{U} \times \bZ_2^{V} \times \bZ_2^{W}
\fe
zero-form symmetry generated by
\ie
	U = \prod_{j \text{ even}} X_j \,, \qquad V = \prod_{j \text{ odd}} X_j \,, \qquad W = \prod_{j} \mathrm{CZ}_{j,j+1} \,,
\fe
where $\mathrm{CZ}_{j,j+1} = \frac{1+Z_j + Z_{j+1} - Z_jZ_{j+1}}{2} = (Z_{j})^{\frac{1-Z_{j+1}}{2}}$ is the Controlled-$Z$ (CZ) gate acting on qubits on sites $j$ and $j+1$. Note that $U$ acts only on the even sites, $V$ acts only on the odd sites, and $W$ acts on all of them.  This is consistent because we took $L$ to be even. The $\bZ_2^{W}$ symmetry acts on local operators as\footnote{In our conventions, $U : \mathcal{O} \mapsto U \mathcal{O} U^\dagger$.}
\ie
	W : X_j \mapsto Z_{j-1} X_j Z_{j+1} \,, \qquad Z_j \mapsto Z_j \,.
\fe

This model is closely related to the Levin-Gu model \cite{Levin:2012yb,Chen:2011bcp,Cheng:2022sgb,Seifnashri:2024dsd}. In particular, our diagonal $\bZ_2$ symmetry generated by $UVW$, is the anomalous symmetry of the Levin-Gu model and is sometimes referred to as the $\mathrm{CZ}X$ symmetry. Also, this model can be seen as a deformation of the example in Subsection \ref{simpleexample}, by breaking the $\bZ_2^{m} \times \U(1)_w \rtimes \bZ_2^{X}$ symmetry there to its subgroup $\bZ_2^{m} \times \bZ_2^{w} \times \bZ_2^{X}$, which is isomorphic to $G_\UV$. In particular, for $h_j = -1$, the Hamiltonian \eqref{lattice.H} describes the compact boson at radius $R=(1/\sqrt{2}) R_\text{self dual}$; see, e.g., \cite{Pace:2024oys}.

\subsection{The UV anomaly}

The $G_\UV$ symmetry of \eqref{GUVlattice} has a ``type-III'' anomaly described by the anomaly theory
\ie
	\alpha_\UV  = \pi i \int \mathcal{A}_U \cup \mathcal{A}_V \cup \mathcal{A}_W \,, \label{type3.anomaly}
\fe
where $\mathcal{A}_U,\mathcal{A}_V,\mathcal{A}_W$ are background gauge fields for the symmetries generated by $U,V,W$. The anomaly \eqref{type3.anomaly} is characterized by the projective action of the symmetry in the presence of various symmetry defects / twisted boundary conditions. Specifically, inserting a defect for any of the three $\bZ_2$ symmetries, the remaining $\bZ_2 \times \bZ_2$ symmetry acts projectively on the defect Hilbert space. (See \cite{Cheng:2022sgb, Seifnashri:2023dpa} for a general discussion of anomalies on the lattice.) Specifically, for the $\bZ^{UVW}\subset \bZ_2^{U} \times \bZ_2^{V} \times \bZ_2^{W}$ symmetry generated by $UVW$, we substitute $\mathcal{A}_U=\mathcal{A}_V=\mathcal{A}_W$ in \eqref{type3.anomaly} and we recover the pure $\bZ_2$ anomaly of \cite{Levin:2012yb}.

We now study the defect Hamiltonians to derive the anomaly \eqref{type3.anomaly}.
Defect Hamiltonians for the three $\bZ_2$ symmetries are given by
\ie
	H_U^{J_\oo} &= - \sum_{j \neq J_\oo} h_j \left( X_j + Z_{j-1} X_j Z_{j+1} \right) - h_{J_\oo} X_{J_\oo}(1 - Z_{J_\oo-1}Z_{J_\oo+1}) \,, \\
	H_V^{J_\ee} &= - \sum_{j \neq J_\ee} h_j \left( X_j + Z_{j-1} X_j Z_{j+1} \right) - h_{J_\ee} X_{J_\ee}(1 - Z_{J_\ee -1}Z_{J_\ee +1}) \,, \\
	H_W^{J_\ee} &= - \sum_{j \neq J_\ee} h_j \left( X_j + Z_{j-1} X_j Z_{j+1} \right) - h_{J_\ee} X_{J_\ee}(Z_{J_\ee-1} + Z_{J_\ee+1}) \,, \label{defect.H}
\fe
where $J_\ee$ ($J_\oo$) is an even (odd) site of the lattice. In general, $H^J_{D}$ denotes the Hamiltonian of a $D$ defect at site $J$. Note that the defect associated with the symmetry $U$ ($V$) acting on the even (odd) sites is on the odd (even) sites. The defect for $W$ could be either on the even or on the odd sites. For simplicity, we focus on inserting it on even sites.

The original symmetry operators $U, V, W$ do not necessarily commute with the defect Hamiltonians and are modified in the presence of defects. For instance, the $\bZ_2^{W}$ symmetry operator $W$ does not commute with $H_U^{J_\oo}$ of \eqref{defect.H}. In the presence of a $\bZ_2^{U}$ defect on an odd site $J_\oo$, $W$ is modified to $W_U = W Z_{J_\oo}$ that commutes with $H_U^{J_\oo}$.	

The symmetry operators in the presence of the defects are given by
\ie
	U_U &= U \,, \qquad &V_U &= V  \,, \qquad &W_U &= W Z_{J_\oo}\,,\\
	U_V &= U \,, \qquad &V_V &= V \,, \qquad &W_V &= W Z_{J_\ee}\,, \\
	U_W &= U \,, \qquad &V_W &= V Z_{J_\ee} \,, \qquad &W_W &= W \,.  \label{lattice.operators}
\fe
(If we place the $W$ defect in \eqref{defect.H} at an odd site, these equations are modified slightly.)

The only nontrivial projective representations correspond to
\ie
	V_U W_U = - W_UV_U\,, \qquad U_V W_V = - W_V U_V \,, \qquad U_W V_W = - V_W U_W \,. \label{uv.proj.alg}
\fe
This verifies that we have the type-III anomaly \eqref{type3.anomaly}.

\subsection{The IR theory}

Let us consider the phase where $h_j=1$ for odd $j$ and $h_j=0$ for even $j$ (or $h_\text{odd} \gg | h_\text{even}|$). The Hamiltonian is
\ie
	H = - \sum_{j \text{ odd}} \left( X_j + Z_{j-1} X_j Z_{j+1} \right) \,, \label{H2}
\fe
which is exactly solvable. The terms in this Hamiltonian commute with each other, and we can minimize each term separately by setting $X_j = 1$ and $Z_{j-1}Z_{j+1}=1$ for odd $j$. The theory is gapped with two ground states given by
\ie
	\ket{\uparrow} = \ket{+0+0 \cdots +0} \qquad \text{and} \qquad \ket{\downarrow} = \ket{+1+1 \cdots +1} \,, \label{g.states}
\fe
where $Z_j \ket{z_1z_2 \cdots z_{L-1}z_L} = (-1)^{z_j} \ket{z_1z_2 \cdots z_{L-1}z_L}$ for $z_j \in \{0,1\}$, and $\ket{\pm} = (\ket{0}\pm\ket{1})/\sqrt2$.

In the thermodynamic limit, the $G_\UV$ symmetry of \eqref{GUVlattice} is spontaneously broken to its $\bZ_2^{V} \times \bZ_2^{W}$ subgroup. There are two superselection sectors corresponding to the two ground states \eqref{g.states}. An order parameter for this symmetry breaking is the local operator $Z_{2l}$ on even site $j=2l$. Its expectation value is independent of $j=2l$. As such, it generates a one-form symmetry in the low-energy theory that acts on the two ground states by a sign.

The faithful symmetry of the IR theory is $\bZ_2^{U} \times \bZ_2^{B,(1)}$. It is generated by\footnote{In the continuum field theory language, we have the 1+1d $\bZ_2$ gauge theory described by the action $S = \frac{2}{2\pi}\int a \mathrm{d} \phi$ \cite{Maldacena:2001ss,Banks:2010zn,Kapustin:2014gua}. The Wilson line $e^{i\int a}$ is identified with the zero-form symmetry operator $U$, and $e^{i\phi}$ is identified with one-form symmetry operator $B$.}
\ie
	U = \prod_{j \text{ even}} X_j \qquad \text{and} \qquad B=Z_{2l} \,. \label{ir.symm.ops}
\fe
Crucially, the two symmetry operators anticommutes with each other:
\ie
	BU=-UB \,. \label{ir.proj.alg}
\fe
This leads to an 't Hooft anomaly described by
\ie
	\alpha_\IR  = \pi i \int \mathcal{A}_U \cup {\cal B}_\IR  \,, \label{lattice.ir.anomaly}
\fe
where ${\cal B}_\IR $ is the background gauge field for the one-form symmetry generated by the symmetry operator $B$.

\subsection{The relation between the UV and IR and symmetry transmutation}

We now find the relation between the UV and IR symmetry operators by looking at their action on the two ground states \eqref{g.states}. In particular, we are interested in this map in the presence of various defects.

Consider the defect Hamiltonian of $U$ at site $J_\oo = 1$ (set $h_\text{odd}=1$, $h_\text{even}=0$ in \eqref{defect.H})
\ie
	H_U^{1} &= - \sum_{j = 3,5, \cdots, L-1}  \left( X_j + Z_{j-1} X_j Z_{j+1} \right) - X_{1}(1 - Z_{L}Z_{2}) \,.
\fe
This Hamiltonian is not frustration-free. Consequently, its ground state energy is higher than that of the untwisted Hamiltonian $H$ in \eqref{H2}. Therefore, the defect Hilbert space of a single $U$ defect has no state in the low-energy theory. This can be explained by reducing the anomaly \eqref{lattice.ir.anomaly} on a circle with a $U$ defect. The resulting QM theory has the anomaly $\pi i \int {\cal B}_\IR $ and, as in Appendix \ref{app:qm}, its Hilbert space is empty.

This lack of states in the IR $U$-defect Hilbert space should be contrasted with the situation in the UV theory.  There, the $U$-defect Hilbert space does have states.  The IR observer interprets these states as having a non-topological defect charged under the one-form symmetry $B$. This situation is analogous to the one discussed in Section \ref{sec:qed3.massive} (see Figure \ref{circle.with.defect}).

Let us discuss the theory in the presence of $V$ and $W$ defects. Since these two symmetries are not spontaneously broken, their defect Hilbert spaces are non-empty in the IR. Specifically, the defect Hamiltonians of $V$ and $W$ (set $h_\text{odd}=1$, $h_\text{even}=0$ in \eqref{defect.H}) are the same as the untwisted Hamiltonian $H$, and thus have exactly two ground states given in \eqref{g.states}.

To find the relation between the UV and IR symmetry operators, we compute the action of the UV symmetry operators \eqref{lattice.operators} on the low-energy ground states \eqref{g.states}. We find
\ie
	U_V &= U \,, \qquad &V_V &= 1 \,, \qquad &W_V &= B\,, \\
	U_W &= U \,, \qquad &V_W &= B \,, \qquad &W_W &= 1 \,.  \label{lattice.operatorsI}
\fe
These relations can be rewritten as
\ie
	U &= U \,, \qquad &V &= B^{w_1} \,, \qquad &W &= B^{v_1}\,, \label{uv.ir.symm.op}
\fe
where $v_1, w_1 \in \bZ_2$ are the holonomies of the background $\bZ_2^{V}$ and $\bZ_2^{W}$ gauge fields $\mathcal{A}_V$ and $\mathcal{A}_W$ in space. These relations nicely match the projective algebra involving UV and IR symmetry operators given in \eqref{uv.proj.alg} and \eqref{ir.proj.alg}.

Note that $V$ and $W$ are symmetry operators of zero-form symmetries and are given by products of local operators over all of space.  On the other hand, as a symmetry operator of a one-form symmetry, $B$ acts at a point and is independent of the location of that point. They are related by \eqref{uv.ir.symm.op}.  Intuitively, the exponents $w_1$ and $v_1$, which are sums over space, account for that fact.

Finally, the relation \eqref{uv.ir.symm.op} can be summarized as the following relation between the UV background gauge fields $\mathcal{A}_U, \mathcal{A}_V, \mathcal{A}_W$ and IR background gauge fields:
\ie
	{\cal B}_\IR  = \mathcal{A}_V \cup \mathcal{A}_W \,. \label{uv.ir.background}
\fe
Compare with \eqref{map.Phi}.

The relations \eqref{uv.ir.symm.op} and \eqref{uv.ir.background} have the following geometric interpretation. The one-form symmetry operator $B$ is the intersection of $V$ and $W$ defects/operators, which intersect at a point. Specifically, $V = B^{w_1}$ corresponds to the intersection of a $\bZ_2^V$ symmetry operator with a $\bZ_2^W$ defect. Similarly, $W = B^{v_1}$ corresponds to the intersection of a $\bZ_2^{W}$ symmetry operator with a $\bZ_2^{V}$ defect. This is similar to the situation in Figure \ref{fig:intersection.symm.lines} after changing $\bZ_2^{X} \times \bZ_2^{Y} \to \bZ_2^{V} \times \bZ_2^{W}$.  This is to be contrasted with the situation with the broken $\bZ_2^{U}$ symmetry. Since it is spontaneously broken, it acts faithfully in the IR ground states of the finite volume theory.

\section{Two-flavor scalar QED$_3$\label{O3d3}}

In this section, we consider the 2+1d version of the theory in Section \ref{sec:2d.u1}.  This is a $\U(1)_\mathrm{EM}$ gauge theory coupled to two charge-one scalar fields $\phi^1$ and $\phi^2$.  The Lagrangian is
\ie\label{CP1UVL}
	\mathcal{L}_\mathrm{Euclidean} = \frac{1}{4e^2} f_{\mu\nu}f^{\mu\nu}-i \frac{k}{4\pi} \varepsilon^{\mu\nu\rho} a_\mu \partial_\nu a_\rho + | D_\mu \phi^i |^2 + m^2 |\phi^i|^2 + V(|\phi^i|^2) \,,
\fe
Here, $f = \mathrm{d} a$ is the field strength of the U(1)$_\mathrm{EM}$ gauge field $a = a_\mu \mathrm{d} x^\mu$. We use the same normalization for the U(1)$_\mathrm{EM}$ gauge field as in Section \ref{sec:2d.u1}. Therefore, the Chern-Simons level $k \in 2\bZ$ is an even integer since we are considering a bosonic theory.

This model and closely related ones have been studied by many authors both on the lattice and in the continuum, exposing many interesting gapped and gapless phases.  See, in particular, the nice review \cite{Sachdev:2010un}.  Here, we will limit ourselves to the continuum theory.  It has an exact magnetic zero-form symmetry, which is absent in some of the lattice models and can also include the Chern-Simons term.  In order to have access to all the phases that are present in the lattice models, we will add various scalar fields and appropriate interactions.

\subsection{The global symmetry and its anomaly}\label{so3globals}

The global symmetry of this model is
\ie
G_\UV= \SO(3)_\mathrm{f} \times \O(2) = \SO(3)_\mathrm{f} \times \left(  \U(1)_\mathrm{m} \rtimes \bZ^C_2 \right) \,.
\fe
As in 1+1d (Section \ref{sec:2d.u1}), $\SO(3)_\mathrm{f}$ is a flavor symmetry acting on the scalar fields and $\bZ^C_2$ is a charge conjugation symmetry that commutes with $\mathrm{SO}(3)_\mathrm{f}$. Unlike the 1+1d problem, here we have a magnetic  $\U(1)_\mathrm{m}$ that acts on monopole operators. Its conserved current is $f={\mathrm{d} a\over 2\pi}$, and the symmetry operator is
\ie
	U_\mathrm{m} (\Sigma, \alpha) = e^{i \alpha \int_\Sigma \frac{\mathrm{d} a}{2\pi}} \,. \label{mag.symm.op}
\fe

When $k \neq 0$, the monopole operator is not gauge-invariant and has charge $k$ under U(1)$_\mathrm{EM}$.\footnote{ Note that U(1)$_\mathrm{EM}$ is the gauge group associated with the dynamical gauge field $a$ and should not be confused with the global symmetry group $\U(1)_\mathrm{m}$ associated with the background gauge field $A_\mathrm{m}$. \label{avsAM}}  A gauge-invariant operator that transforms nontrivially under $\U(1)_\mathrm{m}$ is a dressed monopole operator, which can be chosen to be an SO(3)$_\mathrm{f}$ singlet since $k$ is even. {Specifically, monopole operators are dressed with operators such as $(\varepsilon_{ij} \phi^i \partial_\mu \phi^j)^{\frac k2}$.
Even though for even $k$, the Chern-Simons term does not change the internal symmetry of the theory, it does break its parity/time-reversal symmetry.

The matter fields transform under the extended group (see \eqref{extended.group}):
\ie
\U(2)\rtimes \bZ_2^C =\frac{ \U(1)_\mathrm{EM}  \rtimes ( \SU(2) \times \bZ_2^C ) }{\bZ_2}  \,. \label{qed.extended.group}
\fe
As usual, the $\bZ_2$ quotient means that the matter fields transform as a projective representation of the global symmetry group $\SO(3)_\mathrm{f} \times \bZ_2^C = \frac{ \SU(2) \times \bZ_2^C  }{\bZ_2} $.\footnote{This discussion ignores the monopole operator.} This leads to the relation \eqref{bundle.relation} between U(1)$_\mathrm{EM}$ and $\SO(3)_\mathrm{f} \times \bZ_2^C$ bundles, which is
\ie
	\left(2\int \frac{\mathrm{d}a}{2\pi}\right) \mod{2}= \int \left( w_2[\mathrm{SO}(3)_\mathrm{f}] + C^2 \right)  \,. \label{2+1d.bundle.relation}
\fe

Ignoring the charge conjugation symmetry, the 't Hooft anomaly of the $\SO(3)_\mathrm{f} \times \U(1)_\mathrm{m}$ symmetry is described by
\ie
	\alpha_\UV [ \mathrm{SO}(3)_\mathrm{f} \times \U(1)_{\mathrm m} ] = \pi i \int  w_2[\SO(3)_\mathrm{f}]\cup\left[ \frac{\mathrm{d} A_\mathrm{m}}{2\pi}\right]_2 \,, \label{uv.anomaly.no.c}
\fe
where, $A_\mathrm{m}$ is the background gauge field for the $\U(1)_\mathrm{m}$ magnetic symmetry (see, footnote \ref{avsAM}). Restoring the charge conjugation symmetry, the full anomaly has the more complicated form \cite{Metlitski:2017fmd} (see also \cite{Komargodski:2017smk})
\ie\label{O2ano}
	\alpha_\UV [ \mathrm{SO}(3)_\mathrm{f} \times \O(2) ] = \pi i \int \Big( w_2[\SO(3)_\mathrm{f}] + C^2 \Big) \cup w_2[\O(2)] \,,
\fe	
where $w_2[\O(2)]$ is the second Stiefel-Whitney class of the $O(2)=\U(1)_\mathrm{m} \times \bZ_2^C$ bundle. Restriction of the anomaly to the $\SO(3)_\mathrm{f} \times \bZ_2^\mathrm{m} \times \bZ_2^C \subset \SO(3)_\mathrm{f} \times \U(1)_\mathrm{m} \rtimes \bZ_2^C$ subgroup is
\ie
	\alpha_\UV  \left[ \SO(3)_\mathrm{f} \times \bZ_2^\mathrm{m} \times \bZ_2^C \right]  = \pi i \int \Big( w_2[\SO(3)_\mathrm{f}] + C^2 \Big) \cup \Big( \mathcal{A} \cup C + \mathcal{A}^2 \Big) \,,
\fe
where $\mathcal{A}$ is the background gauge field for $\bZ_2^\mathrm{m} \subset \U(1)_\mathrm{m}$ subgroup and its periods are integers modulo 2.

The expression for the anomaly \eqref{O2ano} is similar to the anomaly in the 1+1d problem \eqref{qed2.anomaly}.  The main difference between them is that in 1+1d, instead of O(2), we had only $\bZ_2^C\subset$ O(2).  We can make these two discussions look closer if we interpret the $\theta$-parameter of the 1+1d problem as a background gauge field for a ($-1$)-form symmetry \cite{Gaiotto:2014kfa,Cordova:2019jnf,Cordova:2019uob,Ferromagnets,Seiberg:2024yig}.  Then, the U(1)$_\mathrm{m}$ part of the 2+1d anomaly \eqref{O2ano} is analogous to the anomaly in the space of coupling constants \eqref{qed2t.anomaly}.

\subsection{Higgs/$\mathbb{CP}^1$ phase -- $\SO(3)_\mathrm{f} \times \O(2)\to \O(2) \times \U(1)_\mathrm{m}$} \label{subsec:higgs}

We first study the $\mathbb{CP}^1$ phase of the theory, where the mass of the scalars is very large and negative, i.e., $m^2<0$ and $|m^2| \gg e^4$. For simplicity, we set the Chern-Simons level $k$ to zero. Here, the gauge group ${\U(1)}_\mathrm{EM}$ is completely Higgsed,  and the $\SO(3)_\mathrm{f}\times \bZ_2^C$ symmetry is spontaneously broken to $\O(2)$.

The IR theory is described by a $\mathbb{CP}^1 = \frac{\SU(2)}{\U(1)}$ non-linear sigma model.  The full UV global symmetry $\SO(3)_\mathrm{f} \times \O(2)$ acts faithfully in the low-energy theory with the same anomaly. The $\O(3) = \SO(3)_\mathrm{f} \times \bZ_2^C$ acts on the $\mathbb{CP}^1$ sigma model coordinates, and $\U(1)_\mathrm{m}$ is realized as the Skyrmion symmetry of the sigma model.

There is no nontrivial relation between the UV and IR symmetry and their anomalies since $G_\UV  = G_\IR $ and $\alpha_\UV  = \alpha_\IR $.

\subsection{Coulomb/$S^1$ phase -- $\SO(3)_\mathrm{f} \times \O(2)\to \SO(3)_\mathrm{f} \times \bZ_2^C$ \label{sec:coloumb}}

Now, we consider the phase with $k=0$, $m^2>0$, and $m^2 \gg e^4$. We can integrate out the heavy scalars and find a pure $\U(1)_\mathrm{EM}$ gauge theory, which can be dualized to a massless compact scalar.\footnote{This scalar leads to a logarithmic confining potential.  Still, we will refer to this phase as a Coulomb phase because it is gapless.} The latter is the Goldstone boson of the spontaneously broken magnetic $\U(1)_\mathrm{m}$ symmetry.}

The faithful symmetry of the low-energy Maxwell theory is
\ie
	G_\IR  = \left( \U(1)_\mathrm{e}^{(1)} \times \U(1)_\mathrm{m} \right) \rtimes \bZ_2^C \,,
\fe
where $\U(1)_\mathrm{e}^{(1)}$ is the electric one-form symmetry generated by the conserved current $J_\mathrm{e} = \frac{1}{e^2} \star \mathrm{d} a$. In terms of the dual compact scalar, this is the winding symmetry. The 't Hooft anomaly of the IR symmetry, ignoring the charge conjugation symmetry, is
\ie
	\alpha_\IR  = 2\pi i \int \frac{B_\IR }{2\pi} \wedge \frac{\mathrm{d} A_\mathrm{m}}{2\pi}\,, \label{2.2.2.IR.anomaly}
\fe
where $B_\IR $ is the background gauge field for the ${\U(1)}_\mathrm{e}^{(1)}$ symmetry, and it is normalized such that its periods, $\int B_\IR$, are defined modulo $2\pi$.

\subsubsection*{Symmetry transmutation}

The UV symmetry $\SO(3)_\mathrm{f}$ does not act faithfully on the local operators of the IR theory and is transmuted into $\bZ_2^{(1)}\subset\U(1)_\mathrm{e}^{(1)}$ one-form symmetry. Indeed, from the relation \eqref{2+1d.bundle.relation} and $\frac{1}{2\pi} \int B_\IR  = \left(\int \frac{\mathrm{d} a}{2\pi} \right)\mod{1}$, we find
\ie
	{1\over \pi}\left(\int B_\IR\right) \mod 2 =  \int \Big( w_2[\SO(3)_\mathrm{f}] + C^2 \Big) \,. \label{coloumb.background.relation}
\fe
In other words, nontrivial background gauge fields for the UV zero-form symmetry leads to a nontrivial background gauge field for the IR one-form symmetry through the relation above. Note that we can only turn on a $\bZ_2$ one-form symmetry background gauge field since the right-hand side of \eqref{coloumb.background.relation} has holonomies equal to $0$ or $1/2$ mod $1$. Therefore, we have symmetry transmutation from $\O(3) = \SO(3)_\mathrm{f} \times \bZ_2^C$ into the $\bZ_{2,\mathrm{e}}^{(1)}$ subgroup of the IR one-form symmetry ${\U(1)}_\mathrm{e}^{(1)}$.

Substituting the relation \eqref{coloumb.background.relation} in the IR anomaly \eqref{2.2.2.IR.anomaly}, and ignoring the charge conjugation symmetry, we see that the anomaly of the UV theory \eqref{uv.anomaly.no.c}, is matched in the IR
\ie
	\alpha_\UV  \left[\SO(3)_\mathrm{f} \times \U(1)_\mathrm{m} \right] = \pi i \int  w_2[\SO(3)_\mathrm{f}] \cup\left[\frac{\mathrm{d} A_\mathrm{m}}{2\pi}\right]_2 \,.
\fe

\subsection{SO(3) group manifold sigma model \label{sec:so3.sigma}}

Here we discuss a phase in which the $\SO(3)_\mathrm{f}$ global symmetry is completely broken.  This phase can be obtained by the following deformation of the UV theory.

For reasons that will be clear soon, we add to the UV theory two neutral scalar fields, $\vec{n}_a(x)$ (with $a=1,2$) that transform as a vector of $\SO(3)_\mathrm{f}$ symmetry. We couple them to the original scalar fields in \eqref{CP1UVL} via the coupling
\ie\label{nphicou}
	\delta \mathcal{L} = \sum_{a=1,2} g_a \vec{n}_a\cdot (\phi^i)^*   \vec{\sigma}_{ij} \phi^j  \,,
\fe
such that the global symmetry of the system remains $\SO(3)_\mathrm{f} \times \O(2)$, and the charge conjugation symmetry $C$ acts as $C : \vec{n}_{a}(x) \mapsto - \vec{n}_{a}(x)$.

When the scalars, $\vec{n}_a(x)$, have a large and positive mass square, they decouple and we recover the model without them.

Now, consider a phase where the original scalars $\phi^i$ have a large and positive mass square, but $\vec{n}_a$ condense. Without loss of generality, we can parameterize the leading order terms in  the potential for $\vec n_a$ as
\ie
	V(\vec{n}_a) = \lambda_1 \Big( \vec{n}_1 \cdot \vec{n}_1 - r_1^2 \Big)^2 + \lambda_2 \Big( \vec{n}_2 \cdot \vec{n}_2 - r_2^2 \Big)^2 + \lambda_{12} \Big( \vec{n}_1 \cdot \vec{n}_2 - r_1r_2 \cos(\beta) \Big)^2 \,,
\fe
where $r_1 \neq r_2$ are positive real numbers, and $\beta \neq 0 ,\pi$. The minimum of the potential is
\ie
	\langle \vec{n}_1 \rangle = \langle \Phi \rangle \begin{pmatrix}
r_1 \\
0 \\
0
\end{pmatrix}  \qquad \text{and} \qquad \langle \vec{n}_2 \rangle = \langle \Phi \rangle \begin{pmatrix}
r_2 \cos(\beta) \\
r_2 \sin(\beta) \\
0
\end{pmatrix}\,,
\fe
where $\Phi$ is an $\SO(3)$-valued scalar field, i.e., a real and orthogonal $3\times3$ matrix with determinant one.

It is now clear why we needed the two scalars $\vec n_a$.  The expectation value of $\vec n_1$ breaks $\O(3)\to \O(2)$ and then the second one breaks it to $\bZ_2$.  In the vacuum where $\langle \Phi\rangle$ is the unit matrix, the unbroken $\bZ_2$ flips the sign of $n_a^3$.

Using a canonically normalized kinetic terms for $\vec n_a$, i.e., ${1\over 2}(\partial_\mu\vec n_a)^2$, we find that the low-energy theory is an $\SO(3)$ sigma model described by $\Phi(x)$ with the Lagrangian
\ie
	\mathcal{L} ={1\over 2} \Tr \left[ \partial_\mu \Phi \begin{pmatrix}
r_1^2 + r_2^2 \cos^2(\beta) & r_2^2 \cos(\beta)\sin(\beta) & 0\\
r_2^2 \cos(\beta)\sin(\beta) & r_2^2 \sin^2(\beta) & 0\\
0 & 0 & 0
\end{pmatrix} \partial^\mu \Phi^\mathrm{T} \right] \,.
\fe
The standard sigma model on the SO(3) group manifold has an $\SO(3)_\mathrm{L} \times \SO(3)_\mathrm{R}$ global symmetry.  Here, it is broken explicitly to $\SO(3)_\mathrm{L} \times \bZ_2^R$, which we identify with the UV symmetry $\SO(3)_\mathrm{f} \times \bZ_2^C$.

The faithful global symmetry of the low-energy theory is
\ie
	G_\IR  = \SO(3)_\mathrm{f} \times \bZ_2^C \times \bZ_{2}^{(1)} \,.
\fe
The $\SO(3)_\mathrm{f}$ symmetry acts as $\Phi\mapsto \mathcal{O} \, \Phi$ for $\mathcal{O} \in \SO(3)$.  And the charge conjugation symmetry acts as\footnote{In the vacuum with $\Phi$ is the unit matrix, the combination of this transformation with the unbroken $\bZ_2$ mentioned above, is the original $\bZ_2^C$ transformation that acts as $\vec n_a\to -\vec n_a$.}
\ie
	\bZ_2^C ~:~ \Phi \mapsto \Phi\begin{pmatrix}
-1 & 0 & 0\\
0 & -1 & 0\\
0 & 0 & +1
\end{pmatrix} \,.
\fe
Since $\pi_1(\SO(3)) = \bZ_2$, there is also a winding one-form symmetry $\bZ_{2}^{(1)}$.
The 't Hooft anomaly of the low-energy theory is
\ie
	\alpha_\IR  = \pi i \int \left( w_2[\SO(3)_\mathrm{f}] +C^2 \right) \cup {\cal B}_\IR  \,. \label{ir.anomaly.so3}
\fe
As usual, ${\cal B}_\IR $ is the background gauge field for $\bZ_{2}^{(1)}$.

To see this anomaly, start with the standard SU(2) sigma model with a global $\SO(4)=\frac{\SU(2)_\mathrm{L} \times \SU(2)_\mathrm{R}}{\bZ_2}$ symmetry.  Then, break the SO(4) symmetry, preserving only the subgroup $\frac{\SU(2)_\mathrm{L} \times \bZ_4^\mathrm{R}}{\bZ_2}$, where $\bZ_4^\mathrm{R}$ generates a $\pi$-rotation along the $z$-axis and acts on the SU(2) fields from the right. Gauging the center subgroup $\bZ_2 \subset \SU(2)_\mathrm{L}$ (or equivalently $\bZ_2 \subset \bZ_4^\mathrm{R}$) of this symmetry, we find our $\SO(3)$ sigma model. Since SU(2)$_\mathrm{L}$ and $\bZ_4^\mathrm{R}$ are nontrivial extensions of $\SO(3)_\mathrm{f}$ by $\bZ_2$, the dual one-form symmetry that we find after gauging, $\bZ_{2}^{(1)}$, has the mixed anomaly \eqref{ir.anomaly.so3} with $\SO(3)_\mathrm{f} \times \bZ_2^C$.\footnote{See \cite{Tachikawa:2017gyf} for a general discussion on the relation between symmetry extensions and anomalies before and after gauging discrete subgroups.}

\subsubsection*{Symmetry transmutation}

The only symmetry of the UV that does not act faithfully in the infrared is the $\U(1)_\mathrm{m}$ magnetic symmetry. This symmetry transmutation into the $\bZ_{2}^{(1)}$ one-form symmetry takes place via the map
\ie
	\int {\cal B}_\IR  = \left(\int \frac{\mathrm{d} A_\mathrm{m}}{2\pi}\right) \mod{2} \,. \label{uv.ir.background.so3}
\fe
Substituting \eqref{uv.ir.background.so3} into the IR anomaly \eqref{ir.anomaly.so3}, we correctly reproduce the UV anomaly \eqref{uv.anomaly.no.c}.

The symmetry transmutation here can also be seen from the projective action of the UV symmetry on some line operators. Consider the Gukov-Witten-type line operators \cite{Gukov:2006jk,Gukov:2008sn} defined by deleting a line $\cal C$ from spacetime and imposing holonomy $\pi $ for the $\U(1)_\mathrm{EM}$ gauge field around it.  This can be phrased as having $\pi$-flux along that line $\cal C$, i.e., $ f=\pi \delta({\cal C})$. This means that the line has half-integer $\U(1)_\mathrm{m}$ charge. For the precise relation between symmetry transmutation and the projective action of the UV symmetry on line operators, see Section \ref{sec:general}.

\subsection{TQFT phase -- $\bZ_2$ gauge theory \label{subsec:z2}}

Another phase can be found by adding to the original model a massive complex scalar field $h$ with charge $2$ under $\U(1)_\mathrm{EM}$ with the potential
\ie
	V(h,\bar{h}) = (m_h)^2 |h|^2 + \lambda_h |h|^4 \,.
\fe
It is important that the global symmetry of the model is the same as it was without $h$.  In particular, all the fields with half-integer SU(2) representations have odd $\U(1)_\mathrm{EM}$ charges and all the fields with integer SU(2) representations have even $\U(1)_\mathrm{EM}$ charges.  Therefore, the global symmetry includes $\mathrm{SO}(3)_\mathrm{f}$, as before.

When $n_h^2$ is positive and $(m_h)^2 \gg e^4$ the added scalar does not affect the dynamics.

Instead, consider the phase where $m_h^2$ is negative and $|m_h^2| ,m^2 \gg e^4$. Then $h$ acts as a Higgs field and Higgses $\U(1)_\mathrm{EM}\to \bZ_2$. The low-energy theory is weakly coupled and is given by a pure $\bZ_2$ gauge theory.  It can be described by two $\U(1)$ gauge fields.  These are the microscopic $\U(1)$ gauge field $a$ and the dual of the phase of $h$, $b$ with the action  $S = \frac{2}{2\pi}\int  a \, \mathrm{d} b$. Its faithful global symmetry is
\ie
	G_\IR  = \bZ_{2,\mathrm{e}}^{(1)} \times \bZ_{2,\mathrm{m}}^{(1)} \,,
\fe
where $\bZ_{2,\mathrm{e}}^{(1)}$ ($\bZ_{2,\mathrm{m}}^{(1)}$) is an electric (magnetic) $\bZ_2$ one-form symmetry generated by the magnetic (electric) line operator $e^{i\int b}$ ($e^{i\int a}$).\footnote{The IR theory also has an e-m exchange zero-form $\bZ_2$ symmetry that exchanges the electric and magnetic lines operators/anyons. We ignore this symmetry, since it is emergent and does not play a role in our discussion.}
This one-form symmetry of the low-energy theory has an 't Hooft anomaly
\ie
	\alpha_\IR  = \pi i \int {\cal B}_\mathrm{e} \cup {\cal B}_\mathrm{m} \,, \label{ir.anomaly.z2}
\fe
where ${\cal B}_\mathrm{e}$ and ${\cal B}_\mathrm{m}$ are background gauge fields for the $\bZ_{2,\mathrm{e}}^{(1)}$ and $\bZ_{2,\mathrm{m}}^{(1)}$ one-form symmetry.

\subsubsection*{Symmetry transmutation}

The infrared theory has no local operators. Therefore, the UV zero-form symmetries act trivially on local operators of the IR theory.  Hence, in order to match the UV anomaly, the UV zero-form symmetries must be transmuted into the one-form symmetries of the IR theory. Indeed the UV $\SO(3)_\mathrm{f} \times O(2)$ symmetry is transmuted into $G_\IR $ via the relation
\ie
	{\cal B}_\mathrm{e} = w_2[\SO(3)_\mathrm{f}] ~ , \qquad {\cal B}_\mathrm{m} = \left[\frac{\mathrm{d} A_\mathrm{m} }{2\pi}\right]_2 \,. \label{uv.ir.background.z2}
\fe
For simplicity, we have ignored the charge conjugation symmetry above. We have already encountered the first relation (e.g., in \eqref{uv.ir.map}), which follows from \eqref{2+1d.bundle.relation}. The second relation follows from the fact that $e^{i\int a}$ is the generator of the magnetic one-form symmetry, and therefore, it couples to ${\cal B}_\mathrm{m}$. Moreover, the gauge field $a$ couples to $\frac{\mathrm{d} A_\mathrm{m} }{2\pi}$ in the UV Lagrangian and therefore ${\cal B}_\mathrm{m} = {[\frac{\mathrm{d} A_\mathrm{m} }{2\pi}]}_2$.\footnote{Note that the IR background field ${\cal B}_\mathrm{m} = \left[\frac{\mathrm{d} A_\mathrm{m} }{2\pi}\right]_2$ for $\bZ_{2,\mathrm{m}}^{(1)}$ is the same as the IR background field ${\cal B}_\IR $ in \eqref{uv.ir.background.so3}.  Yet, in Section \ref{sec:so3.sigma} the corresponding symmetry arises as a winding symmetry in SO(3), while here it is a magnetic symmetry.}

In summary, the UV global symmetries $\SO(3)_\mathrm{f}$ and $\U(1)_\mathrm{m}$, respectively, are transmuted into the $\bZ_{2,\mathrm{e}}^{(1)}$ and $\bZ_{2,\mathrm{m}}^{(1)}$ one-form symmetries of the low-energy theory. Substituting \eqref{uv.ir.background.z2} into the IR anomaly \eqref{ir.anomaly.z2} matches the UV anomaly \eqref{uv.anomaly.no.c}.

We would like to end the discussion of this $\bZ_2$ gauge theory phase with an interesting lesson.  It used to be said that if the UV symmetry has nonzero 't Hooft anomalies, then either the symmetry is spontaneously broken or the IR theory is gapless.  Later, gapped phases matching anomalies in unbroken $\U(1)$ and time-reversal were found in \cite{Vishwanath:2012tq,Metlitski:2013fal,Bonderson:2013pla ,Chen:2013jha,Wang:2013uky,Wang:2013zja, Metlitski:2015eka, Mross:2014gla,Seiberg:2016rsg}.  Although in these examples, the $\U(1)$ symmetry does not act faithfully in the low-energy theory, the time-reversal symmetry acts there as an ordinary time-reversal symmetry.  (See the discussion in Section \ref{necessary conditions}.)  The interesting aspect of this $\bZ_2$ gauge theory phase is that the entire anomalous UV symmetry is unbroken and does not act faithfully in the IR.  Yet, its anomalies are matched in the IR via symmetry transmutation.

\subsection{TQFT phase -- U(1)$_k$ CS \label{subsec:u1k.cs}}

Here, we restore the U(1)$_k$ Chern-Simons (CS) term.  We take $k\in 2\bZ$, such that the theory is bosonic. For $m^2 \gg e^4$, the matter fields are irrelevant, and then the low-energy theory is described by a pure $\U(1)_k$ Chern-Simons theory with the Euclidean Lagrangian
\ie\label{CP1UVLL}
- i\frac{k}{4\pi} a\wedge\mathrm{d}a -  i\frac{1}{2\pi} a \wedge \mathrm{d} A_\mathrm{m}\,,
\fe
where we included the coupling to the background gauge field $ A_\mathrm{m}$ in the UV theory.

As in \eqref{2+1d.bundle.relation}, $a$ is not a standardly normalized gauge fields.  Depending on the UV background fields, its fluxes can be half-integers
\ie
	\left(2\int \frac{\mathrm{d}a}{2\pi}\right) \mod{2}= \int  w_2[\mathrm{SO}(3)_\mathrm{f}]   \,, \label{2+1d.bundle.relationk}
\fe
where, for simplicity, we have set $C=0$.

The faithful global symmetry of this theory is\footnote{There may exist additional emergent quantum symmetries associated with distinct prime factors of $k/2$, as discussed, for example, in \cite{Delmastro:2019vnj}. We ignore such emergent symmetries here.}
\ie
	G_\IR  = \bZ_k^{(1)} \rtimes \bZ_2^C\,,
\fe
where $\bZ_k^{(1)}$ is a one-form symmetry.  The periods of its background gauge field ${\cal B}_\IR$ are integers modulo $k$.

To find the effect of the background field, ${\cal B}_\IR$, it is convenient to first redefine
\ie
\hat a=a+ {1\over k} A_\mathrm{m}\,.
\fe
The unusual factor of $1\over k$ means that $\hat a$ is not a standard $\U(1)$ gauge field and its fluxes can be fractional.  We will soon return to that.  In terms of $\hat a$, the Lagrangian \eqref{CP1UVLL} is
\ie\label{CP1UVLLs}
 -i\frac{k}{4\pi} \hat a \wedge \mathrm{d}\hat a +  i\frac{1}{4\pi k}  A_\mathrm{m} \wedge \mathrm{d} A_\mathrm{m}\,.
\fe
Ignoring for the moment the ill-defined second term, we have a standard $\U(1)_k$ Chern-Simons theory, except that the fluxes of the dynamical gauge field $\hat a$ are constrained to satisfy
\ie
	\left(k\int \frac{\mathrm{d}\hat a}{2\pi}\right) \mod{k}= \int \left( {k\over 2} w_2[\mathrm{SO}(3)_\mathrm{f}] +\left[{\mathrm{d} A_\mathrm{m} \over 2\pi}\right]_k\right) \,. \label{2+1d.bundle.relationks}
\fe
(This is well-defined because $k$ is even.)
These fluxes can be fractional, where the fractional part is determined by the background fields.

We find the same fractional fluxes when we couple the system to a background field for its one-form symmetry $\bZ_k^{(1)}$
\ie	
	{\cal B}_{\IR } =   \frac{k}{2} w_2[\mathrm{SO}(3)_\mathrm{f}]  + \left[\frac{\mathrm{d}A_\mathrm{m}}{2\pi}\right]_k\,. \label{uv.ir.background.relationk}
\fe
We interpret this equation as the symmetry transmutation map $\Phi$ from the UV background fields to the IR field ${\cal B}_\IR$.

On general grounds, $\U(1)_k$ coupled to a background gauge field ${\cal B}_\IR$ for its $\bZ_k^{(1)}$ symmetry has an 't Hooft anomaly, with the anomaly theory \cite{Kapustin:2014gua,Gaiotto:2014kfa,Hsin:2018vcg}
\ie
	\alpha_\IR  = 2\pi i \int \frac{1}{2k} \mathcal{P}\left({\cal B}_{\IR }\right) \,. \label{u1.k.anomaly}
\fe
Here, $\mathcal{P}\left({\cal B}_{\IR }\right) $ is the Pontryagin square.  It is roughly the square of ${\cal B}_{\IR }$ in the sense that it satisfies $\mathcal{P}({\cal B}_1+{\cal B}_2) = \mathcal{P}({\cal B}_1)+\mathcal{P}({\cal B}_2) + 2 {\cal B}_1 \cup {\cal B}_2$ and $\mathcal{P}(m {\cal B}) = m^2 \mathcal{P}(B)$ for $m \in \bZ$. Importantly, $ \mathcal{P}\left({\cal B}_{\IR }\right)$ is $2k$-periodic.

Next, we check the 't Hooft anomaly matching.  Substituting the symmetry transmutation map \eqref{uv.ir.background.relationk} in the IR anomaly \eqref{u1.k.anomaly}, and restoring the term we ignored in \eqref{CP1UVLLs}, we find
\ie
	2\pi i \int \left( \frac{k}{8} \mathcal{P}\left( w_2[\SO(3)_\mathrm{f}] \right) +  \frac{1}{2} w_2[\SO(3)_\mathrm{f}] \cup \left[  \frac{\mathrm{d}A_\mathrm{m}}{2\pi}\right]_k \right) \, \sim \,  \pi i \int w_2[\SO(3)_\mathrm{f}] \cup \left[ \frac{\mathrm{d}A_\mathrm{m}}{2\pi} \right]_2  \,. \label{U1kIRan}
\fe
The first term above can be expressed in terms of the instanton number of the SO(3)$_\mathrm{f}$ symmetry.  In that form, it is locally well-defined and does not contribute to the anomaly.  (Equivalently, once written in terms of instanton densities, the coefficients of the first term can be continuously set to zero and therefore it do not contribute to the anomaly.) For this reason, we dropped it in the right-hand side of the equation.

We conclude that the symmetry transmutation map \eqref{uv.ir.background.relationk} makes the UV anomaly \eqref{uv.anomaly.no.c} match the IR anomaly \eqref{u1.k.anomaly}.

\section{One-flavor QCD$_4$}\label{sec:QCD}

In this Section, we discuss different phases of one-flavor QCD$_4$ differing by the action of the $\U(1)_{B}$ baryon symmetry in the infrared.

More specifically, we consider $\SU(3)_\mathrm{c}$ gauge theory in 3+1d coupled to a single Dirac quark $\Psi(x) = (\psi^1(x),\psi^2(x),\psi^3(x))^\mathrm{T}$ transforming in the fundamental representation of $\SU(3)_\mathrm{c}$. The Lagrangian (in Euclidean signature) is\footnote{We use the common normalization of the trace, where $\Tr (T_a T_b) = \frac12 \delta_{a,b}$.}
\ie
	\mathcal{L}_\mathrm{QCD} = \frac{1}{2g^2} \mathrm{Tr} \left( f_{\mu\nu} f^{\mu\nu} \right) +  i \bar{\Psi} \, \slashed{D}_{a+\tilde{A}\mathbbm{1}_{3\times3}} \,\Psi + m \bar \Psi \Psi -i \frac{\theta}{16\pi^2} \mathrm{Tr} \left( f_{\mu\nu} \star f^{\mu\nu}  \right) \,, \label{L.QCD}
\fe
where $f_{\mu\nu} = \partial_\mu a_\nu - \partial_\nu a_\mu -i [a_\mu, a_\nu]$ is the $\SU(3)_\mathrm{c}$ field strength, and $\star f^{\mu\nu} = \frac{1}{2}\varepsilon^{\mu\nu\rho\lambda} f_{\rho\lambda}$. As we will explain below, $\tilde{A}$ that appears in the covariant derivative of the quark field is related to the background gauge field for the global baryon number symmetry $\U(1)_{B}$.

\subsection{The global symmetries}

The global symmetry that we are interested in is the baryon number symmetry $\U(1)_{B}$ that acts on the quarks. We normalize it such that gauge-invariant operators have integer baryon charges. Therefore, the quarks have charge $B=1/3$.

The quark field $\Psi(x) = (\psi_1(x),\psi_2(x),\psi_3(x))^\mathrm{T}$ transforms under the group
\ie
	\U(3) = \frac{\SU(3)_\mathrm{c} \times \widetilde{\U}(1)}{\bZ_3} \,, \label{total.group.qcd}
\fe
where the $\SU(3)_\mathrm{c}$ subgroup of it is gauged. (See Section \ref{gaugetheoryexamples}.) Therefore, its coupling through the covariant derivative in \eqref{L.QCD} is to a U(3) gauge field
\ie\label{aAtildeA}
	a + \tilde{A} \, \mathbbm{1}_{3\times3} \,.
\fe
As we said above, $a$ is a dynamical $\mathrm{SU(3)}_\mathrm{c}$ and then we identify
\ie
\text{Tr}\left(a + \tilde{A} \, \mathbbm{1}_{3\times3} \right) = A
\fe
as the background gauge field of the baryon number symmetry $\U(1)_{B} = \frac{\mathrm{\widetilde{U}(1)}}{\bZ_3}$.

Importantly, while the combination $a + \tilde{A} \, \mathbbm{1}_{3\times3}$ is a well-defined U(3) gauge field, its decomposition into an SU(3)$_\mathrm c$ gauge field $a$ and a $\widetilde\U(1)$ gauge field $\tilde A$ is not.

As a consequence of that, depending on the background $A$, the dynamical gauge field $a$ can be in a twisted SU(3)$_\mathrm{c}$ bundle.  Specifically,
\ie
	\left(\int \frac{\mathrm{d}A}{2\pi}\right) \mod{3} = \int u_2[ a ] \,, \label{twisted.bundle}
\fe
where $u_2[a] \in H^2(\mathcal{M} , \bZ_3)$ is the second characteristic class of a PSU(3)$_\mathrm{c}$ gauge bundle.

We conclude that some background fields $A$ force the gauge field $a$ to be in PSU(3)$_\mathrm{c}$ bundles that are not SU(3)$_\mathrm {c}$ bundles, i.e., it is as if the SU(3)$_\mathrm{c}$ gauge field $a$ is twisted by a $\bZ_3^{(1)}$ one-form symmetry.\footnote{Many authors have used similar twisted background gauge fields to force twisted bundles for the dynamical fields in such gauge theories.  One early example is \cite{Cohen:1983cc} and a recent closely related example is \cite{Tanizaki:2022ngt}.}

Below, we will see how we can use these twisted SU(3) bundles to explore the phases of the theory.

\subsection{Confining phase -- heavy quark limit}

We begin with the heavy quark limit, where $m \gg \Lambda_\mathrm{QCD}$. In this limit, below the mass of the quark, we have a symmetry transmutation.

Specifically, at energy scales $E$ with $m \gg E \gg \Lambda_\mathrm{QCD}$, we can integrate out the quarks, and we are left with a pure Yang-Mills (YM) theory. The $\U(1)_{B}$ baryon symmetry does not act faithfully on the local operators of the effective YM theory. As we discussed above, the $\U(1)_{B}$ symmetry is transmuted into the $\bZ_3^{(1)}$ one-form symmetry of the YM theory.

The symmetry transmutation is described by the relation
\ie
	{\cal B}_\IR = \left[\frac{\mathrm{d} A}{2\pi}\right]_3\,,
\fe
that is derived from \eqref{twisted.bundle}, and the fact that ${\cal B}_\IR = u_2[a]$. Thus, even though there is no one-form symmetry in the UV, we can turn on a one-form symmetry background gauge field by activating a UV background gauge field for the baryon number symmetry. Note that in this case, the symmetry transmutation is not needed for anomaly matching.

In the far infrared limit, however, the theory is believed to be trivially gapped without any faithful global symmetry. Therefore, the symmetry transmutation is only visible at energy range $m \gg E \gg \Lambda_\mathrm{QCD}$.

\subsection{Deconfined phase -- adding an adjoint Higgs field}

Here, we discuss the deconfined phase of QCD by adding a scalar (Higgs) field in the 8-dimensional adjoint representation of $\SU(3)_\c$.  We represent it as a hermitian $3\times 3$ traceless matrix $\Phi_8$ and write the potential
\ie
	V(\Phi_8) = m_8^2 |\Phi_8|^2 + \mu_8 \det \Phi_8 + \lambda_8 |\Phi_8|^4 + \cdots \,, \label{V8}
\fe
where $|\Phi_8|^2 = \Tr\Phi_8^2$ and $|\Phi_8|^4 =(\Tr\Phi_8^2)^2$.\footnote{Note that since $\Tr \Phi_8 = 0$, we have $\Tr \Phi_8^3 = 3 \det \Phi_8$ and $\Tr \Phi_8^4 = \frac12 (\Tr \Phi_8^2)^2$ and therefore we did not include them.  The ellipse represent higher order terms that make the minima isolated.}
The scalar field $\Phi_8$ is invariant under the $\U(1)_{B}$ symmetry and does not directly couple to the quarks. We can also add a Yukawa coupling like $\bar \Psi \Phi_8 \Psi$ without changing the symmetry. When $m_8^2 \gg \Lambda_\mathrm{QCD}^2$, the scalar field decouples from the low-energy theory and does not change the physics at long distances.

However, we are interested in the case with negative $m_8^2$ and $|m_8^2|, m^2 \gg \Lambda_\mathrm{QCD}^2$. Then, $\Phi_8$ acts as a Higgs field.  For generic coupling, its expectation value has three distinct eigenvalues and then it Higgses $\SU(3)_\c$ down to its $\bZ_3$ subgroup.  The low-energy theory is weakly coupled and is described by a pure $\bZ_3$ gauge theory. It can be represented by the effective action $S = \int \frac{3}{2\pi} a^{(1)} \mathrm{d} b^{(2)}$, where $a^{(1)}$ is a U(1) one-form and $b^{(2)}$ is a U(1) two-form dynamical gauge field. The faithful global symmetry of the infrared theory is\footnote{We ignore the charge conjugation symmetry both in the UV and IR.}
\ie\label{higherformn}
	G_\IR = \bZ_{3,\ee}^{(1)} \times \bZ_{3,\mm}^{(2)} \,.
\fe
Here, $\bZ_{3,\ee}^{(1)}$ is an electric one-form symmetry acting on the Wilson line $e^{i \int a^{(1)}}$ and $\bZ_{3,\mm}^{(2)}$ is a magnetic two-form symmetry acting on the Gukov-Witten surface operator $e^{i \int b^{(2)}}$.

Because of the relation  \eqref{twisted.bundle} and ${\cal B}_\IR = u_2[a]$, the UV $\U(1)_{B}$ symmetry is transmuted into the electric one-form symmetry of the IR theory via the relation
\ie
	{\cal B}_\IR = \left[\frac{\mathrm{d} A}{2\pi}\right]_3 \,.
\fe
This one-form symmetry is spontaneously broken.

Therefore, the confined and deconfined phases are distinguished by different realizations of the symmetries in the infrared. In the first phase, there is no faithful global symmetry in the infrared, and $\U(1)_{B}$ acts trivially in the deep IR theory. In the deconfined phase, the $\U(1)_{B}$ symmetry is transmuted into a one-form symmetry in the IR and this symmetry is spontaneously broken.\footnote{We emphasize that the new one-form symmetry arising from symmetry transmutation here differs from emergent one-form symmetries, such as those discussed in \cite{Cordova:2022rer,Cherman:2022eml,Cherman:2023xok}.} The authors of \cite{Gagliano:2025gwr} independently discussed this phenomenon from a representation theory perspective.

The standard way of phrasing this result is that in the confining phase the IR theory is trivially gapped, while in this deconfined phase, we have a $\bZ_3$ gauge theory.  Our discussion recognizes the spontaneously broken one-form symmetry in that phase as a result of symmetry transmutation of the exact UV $\U(1)_B$ symmetry.

\subsection{Higgs phase -- adding a Higgs field in the 6 and 8}

Here, we discuss a phase that completely Higgses $\SU(3)_\c$. To do that, in addition to the adjoint scalar field $\Phi_8$, we add a scalar field $\Phi_6$ in the 6-dimensional representation of $\SU(3)_\c$ described by $3\times3$ symmetric complex matrix.\footnote{This model and its time-reversal symmetry will be discussed in \cite{DumitrescuHsinUpcoming} as an example of a nontrivial Higgs-confinement transition.} The scalar fields $\Phi_6$ and $\Phi_8$ transform as
\ie
	\U(3) ~:~ \begin{aligned}
	&\Phi_6 \mapsto \mathcal{U} \, \Phi_6 \, \mathcal{U}^\mathrm{T} \\
	&\Phi_8 \mapsto \mathcal{U} \, \Phi_8 \, \mathcal{U}^\dagger
	\end{aligned} ~,
\fe
where $\mathcal{U}$ is a $\U(3)$ matrix. In particular $\Phi_6$ has charge $B=2/3$ under the $\U(1)_{B}$ symmetry.

We couple $\Phi_6$ and $\Phi_8$ to the quark via a Yukawa coupling
\ie
 y_6 \, \Psi^{\mathrm{T}} \mathcal{C} (\Phi_6)^\dagger \Psi +\mathrm{c}.\mathrm{c}. + y_8 \bar \Psi \Phi_8 \Psi   \,,
\fe
where the matrix $\mathcal{C}$ acts only on the Lorentz indices and satisfies $\mathcal{C} \gamma^\mu = - (\gamma^\mu)^\mathrm{T}\mathcal{C}$.  We also include a scalar potential (compare with \eqref{V8})
\ie
	V(\Phi_6,\Phi_8) &=  m_6^2 |\Phi_6|^2 +m_8^2 |\Phi_8|^2+ \cdots ~. \label{Higgs.potential}
\fe
Here, $|\Phi_6|^2 = \Tr(\Phi_6^\dagger \Phi_6)$.  The ellipses include higher-order terms.

When $m_6^2, m_8^2 \gg \Lambda_\mathrm{QCD}^2$, we have $\langle \Phi_6 \rangle=\langle \Phi_8 \rangle=0$. The scalar fields are massive and decouple at low-energies, and we are left with the original model.

Next, consider the phase where $m_6^2$ and $ m_8^2$ are negative and $|m_6^2|, |m_8^2|, m^2 \gg \Lambda_\mathrm{QCD}^2$.  We set the coefficients of the potential such that $\langle \det\Phi_6 \rangle = 0$, and thus $\U(1)_{B}$ is not spontaneously broken.  (Note that $\det\Phi_6$ has baryon charge $B=2$). Then the expectation value of $\Phi_6$ can be parameterized as
\ie
	\langle \Phi_6 \rangle = \mathcal{U} \begin{pmatrix}
v_1 & 0 & 0\\
0 & v_2 & 0\\
0 & 0 & 0
\end{pmatrix} \mathcal{U}^T \,, \qquad \text{for } \mathcal{U} \in \U(3) \,. \label{VEV}
\fe
We take the potential to lead to distinct, nonzero $v_1 $ and $v_2$, and to distinct nonzero eigenvalues of $\langle \Phi_8 \rangle$ as before.

In this phase, the scalar fields completely Higgs $\SU(3)_\c$ and the global $\U(1)_{B}$ symmetry is unbroken. To see this, we first note that the VEV \eqref{VEV} breaks U(3) down to $\bZ_2 \times \bZ_2 \times \U(1)_{B}$ given by the U(3) transformation
\ie
	\mathcal{U} = \begin{pmatrix}
\epsilon_1 & 0 & 0\\
0 & \epsilon_2 & 0\\
0 & 0 & e^{i\beta} \epsilon_1\epsilon_2
\end{pmatrix} \qquad \text{for } \epsilon_1, \epsilon_2 = \pm1 \,.
\fe
Since $\det \mathcal{U}=e^{i\beta}$, the parameter $\beta$ parameterizes the baryon symmetry transformation.  The parameters  $\epsilon_1, \epsilon_2$ parameterize the unbroken $\bZ_2 \times \bZ_2$ subgroup of $\SU(3)_\c$. Thus $\Phi_6$ alone does not completely Higgs $\SU(3)_\c$.  However, the adjoint scalar $\Phi_8$ Higgses the remaining  $\bZ_2 \times \bZ_2$ gauge symmetry.  (The role of $\Phi_6$ is to Higgs the center of the SU(3) gauge symmetry and the role of $\Phi_8$ is to Higgs this $\bZ_2 \times \bZ_2$.)

In summary, here we have a complete Higgs phase. The low-energy theory is weakly coupled and is trivially gapped. The faithful global symmetry of the infrared theory is trivial. We note that in this Higgs phase, there is no symmetry transmutation, even at some intermediate energy scale. This is because the Wilson lines end on the VEV of the scalar fields and, therefore, violate the necessary conditions for symmetry transmutation. This is to be contrasted with the other two phases, where below the mass of the matter fields we had symmetry transmutation from $\U(1)_{B}$ to $\bZ_3^{(1)}$\,.

\subsection{Summary}

Let us summarize and compare the three phases of one-flavor QCD studied above.
\begin{itemize}
\item \textbf{Confining:} In the confining phase, the $\U(1)_{B}$ symmetry is transmuted into a $\bZ_3^{(1)}$ one-form symmetry at energy scales $m \gg E \gg \Lambda_\mathrm{QCD}$. This phase is believed to be trivially gapped, and thus, the $\bZ_3^{(1)}$ one-form symmetry acts trivially on the confining vacuum.
\item \textbf{Deconfined:} In the de-confining phase, the $\U(1)_{B}$ symmetry is transmuted into a \emph{spontaneously broken} $\bZ_3^{(1)}$ one-form symmetry. The latter acts nontrivially on the IR TQFT described by a $\bZ_3$ gauge theory.
\item \textbf{Higgs:} Finally, in the Higgs phase, there is no symmetry transmutation, and $\U(1)_{B}$ acts completely trivially at energy scales below the Higgs VEV.
\end{itemize}

We observe that the confining and Higgs phases are indistinguishable in the deep IR. This is compatible with the Higgs-confinement continuity lore \cite{Fradkin:1978dv,Banks:1979fi,Raby:1979my,Dimopoulos:1980hn}, which states that the Higgs and confining phases are smoothly connected in the absence of global symmetries. Recently, this lore has been revisited in the presence of additional global symmetries in the UV. In particular, \cite{Verresen:2022mcr,Thorngren:2023ple,Dumitrescu:2023hbe} have pointed out the possibility that the Higgs and confining phases can be in different symmetry-protected topological (SPT) phases. This is sometimes referred to as ``Higgs=SPT.'' These authors studied the system in a regularization scheme where the confining phase has a trivial partition function, and the Higgs phase is in a nontrivial SPT phase described by an invertible TQFT.

It would be interesting to study the implications of symmetry transmutations for the Higgs-confinement continuity and examine its relation to the ``Higgs=SPT'' scenario.

\section{General remarks on symmetry transmutation \label{sec:general}}

In this section, we comment on various aspects of the phenomenon of symmetry transmutation that we encountered in the examples in the previous sections.

\subsection{Necessary conditions for transmutation}

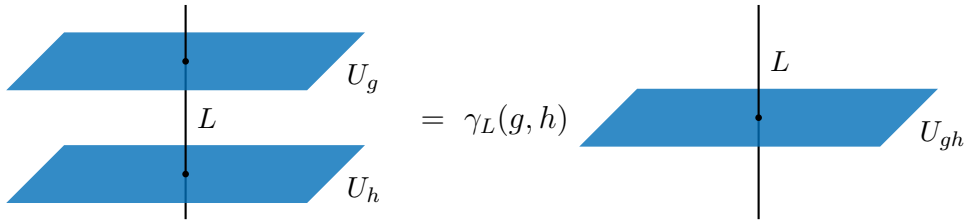
\begin{figure}[ht]
    \centering
    \raisebox{-35pt}{\begin{tikzpicture}[scale=1]
	
	\draw[thick, black] (0,-1.6,-1) -- (0,-1,-1);
        \fill[NavyBlue, opacity=0.8] (-2,-1,0) -- (2,-1,0) -- (2,-1,-2) -- (-2,-1,-2) -- cycle;
        \draw[thick, black] (0,-1,-1) -- (0,0.5,-1) node[pos=0.5, right] {\small $L$};
        \fill[NavyBlue, opacity=0.8] (-2,0.5,0) -- (2,0.5,0) -- (2,0.5,-2) -- (-2,0.5,-2) -- cycle;
	\draw[thick, black] (0,0.5,-1) -- (0,1.25,-1);
	
        \fill[black] (0,0.5,-1) circle (1.25pt);
        \fill[black] (0,-1,-1) circle (1.25pt);

        \node[below right] at (2,-0.9,-1) {\small $U_h$};
        \node[below right] at (2,0.6,-1) {\small $U_g$};
    \end{tikzpicture}} $~=~ \gamma_L(g,h)$
    \raisebox{-35pt}{\begin{tikzpicture}[scale=1]
	
	\draw[thick, black] (0,-1.6,-1) -- (0,-0.25,-1);
        \fill[NavyBlue, opacity=0.8] (-2,-0.25,0) -- (2,-0.25,0) -- (2,-0.25,-2) -- (-2,-0.25,-2) -- cycle;
        \draw[thick, black] (0,-0.25,-1) -- (0,1.25,-1) node[pos=0.5, right] {\small $L$};
	
        \fill[black] (0,-0.25,-1) circle (1.25pt);

        \node[below right] at (2,-0.15,-1) {\small $U_{gh}$};
    \end{tikzpicture}}
    \caption{The action of the UV zero-form symmetry $G^{(0)}$ on an IR line defect $L$.  The symmetry operators $U_g$, for $g \in G$, can act projectively on the $L$-defect Hilbert space. This is labeled by the phase $\gamma_L(g,h)$, which is determined by the IR one-form symmetry via $[\gamma] \in H^2(G,K)$.  Intuitively, the phase $\gamma_L(g,h)$ arises from the region of the intersection of $L$ with the symmetry operators.  It is associated with a contact term there.  Here, we limit ourselves to $U_g$ that do not change the anyon type.  The more general case, involving permutation of lines is mentioned in the text.}
    \label{fig:proj.action.on.lines}
\end{figure}

What are the necessary conditions to find symmetry transmutation?  For simplicity, let us focus on the symmetry transmutation of a zero-form symmetry $G^{(0)}$ in the UV into a one-form symmetry $K^{(1)}$ in the IR. For this to happen, the following necessary conditions should hold:
\begin{enumerate}

\item The UV theory includes the zero-form symmetry $G^{(0)}$ and does not include the one-form symmetry $K^{(1)}$.  Otherwise, it is obvious that there is no symmetry transmutation.

\item The UV symmetry $G^{(0)}$ should not act faithfully in the IR. Furthermore, for the entire symmetry $G^{(0)}$ to be transmuted into $K^{(1)}$, every symmetry operator in $G^{(0)}$ must act non-faithfully in the IR. Below, we will define more carefully what we mean by faithful action of the symmetry.

\item However, we assume that the UV zero-form symmetry $G^{(0)}$ does act nontrivially on at least some of the lines of the IR theory, without permuting them.  See Figure \ref{fig:proj.action.on.lines}.

\end{enumerate}

The above points are necessary conditions for symmetry transmutation of the entire 0-form symmetry $G^{(0)}$ into $K^{(1)}$. More generally, there exist situations where $G^{(0)}$ acts non-faithfully and a quotient of it acts faithfully in the IR. In such cases, only a subgroup of $G^{(0)}$ is transmuted into $K^{(1)}$.

\subsubsection{Comments about faithfulness}\label{necessary conditions}

The second and third conditions above state that in the IR, the UV zero form symmetry $G^{(0)}$ is non-faithful but nontrivial.  This distinction should be clarified.

A 0-form symmetry $G^{(0)}$ is said to act \emph{faithfully} if every nontrivial element of $G^{(0)}$ acts non-trivially on at least one state in the Hilbert space of the theory on some spatial manifold $\Sigma$.\footnote{More generally, a faithful $p$-form symmetry requires the corresponding symmetry operator to be nontrivial for at least one closed codimension-$p$ submanifold in space.}

The simplest case, which is the most common one, is when $\Sigma$ is a sphere.  In that case, a nontrivial action on the Hilbert space on $\Sigma$ is related to a nontrivial action on point operators.

Interestingly, even when $G^{(0)}$ acts trivially on all point operators (e.g., if there are no point operators at all), it could still act nontrivially on line operators.  For example, in 2+1d, a nontrivial action of $G^{(0)}$ on the Hilbert space on $\Sigma=T^2$ corresponds to action on line operators.  A known example is the action of $G^{(0)}$ by permuting line operators.

And even if the action on the Hilbert spaces on the sphere or the torus is trivial, $G^{(0)}$ can still act nontrivially on the Hilbert space on higher-genus surfaces. Such symmetries were discussed in \cite{Kobayashi:2025ykb} and were referred to as soft symmetries. This action of $G^{(0)}$ corresponds to the action on junctions of line operators.

In this discussion, it is important that the symmetry operators act on the point, or the line, or on a more complicated configuration of lines without touching them.  In other words, the faithful action does not involve contact terms.\footnote{In quantum field theory, we mostly limit ourselves to correlation functions of operators at separated points.  Correlation functions at coincident points are controlled by contact terms.}

The second condition above states that the action of $G^{(0)}$ in the IR is non-faithful. This should be contrasted with the situation mentioned in the third condition and is described in Figure \ref{fig:proj.action.on.lines}.  Here, the action of the symmetry on the line is nontrivial, but it is also non-faithful.  It arises when the line pierces the surface, and therefore, it is associated with contact terms.

In the context of 2+1d TQFTs, the authors of \cite{Benini:2018reh} referred to faithful symmetries, which act without such contact terms, as \emph{intrinsic} symmetries. In contrast, they referred to the piercing action of 0-form symmetries that lead to the fractionalization as \emph{extrinsic} data.

As we have emphasized throughout this paper, non-faithful symmetries, where every symmetry operator acts non-faithfully, can still exhibit ’t Hooft anomalies, which must be matched in the IR via symmetry transmutation.

\subsection{The map between the UV and IR background gauge fields \label{sec:central extension}}

As we said above, the symmetry transmutation is described by a map $\Phi$ from the UV background gauged fields to the IR higher-form background gauge fields:
\ie\label{AtoBmap}
	B_\IR  = \Phi (A_\UV ) \,.
\fe
When both the UV and the IR symmetries are zero-form symmetries, the map \eqref{AtoBmap} contains the same information as the standard homomorphism $\varphi: G_\UV  \to G_\IR $.  In the case of symmetry transformation, we do not have such a homomorphism of symmetry groups, but \eqref{AtoBmap} is still valid.  In this sense, \eqref{AtoBmap} generalizes the standard homomorphism of symmetry groups $\varphi$.

For the case of a symmetry transmutation from a 0-form symmetry $G^{(0)}$ in the UV into a 1-form symmetry $K^{(1)}$ in the IR, the map $\Phi$ is described by a central extension of $G$ by $K$\footnote{Recall our notation that $K^{(1)}$ is a one-form global symmetry based on the group $K$.}
\ie
	1 \to K \to \tilde{G} \to G \to 1 \,. \label{extension}
\fe
Such extensions are classified by the cohomology group $H^2(G, K)$. Given any such extension, there is a canonical map from any $G^{(0)}$ gauge field $A_\UV$ to a $K^{(1)}$ gauge field $B_\IR$, which is independent of the spacetime manifold. See Appendix \ref{app:math}, for more details.

The extension \eqref{extension} follows from the projective action of $G$ on the line operators in the theory. For any line operator $L$, we define a function $\gamma_L: G\times G \to U(1)$ that characterizes the action of $G$ on $L$, as illustrated in Figure \ref{fig:proj.action.on.lines}. Importantly, this projective action is determined by the action of the one-form symmetry $K^{(1)}$.\footnote{Following \cite{Barkeshli:2014cna,Benini:2018reh},  the phase $\gamma $ is a function of $ \widehat{K} \times G\times G \to U(1)$, where $\widehat K$ is the Pontryagin dual of $K$, i.e., its representations.  This map is a homomorphism with respect to the first component, i.e., $\gamma_{a}(g_1,g_2)\gamma_{b}(g_1,g_2) = \gamma_{ab}(g_1,g_2)$. This allows us to rewrite the function $\gamma$ as $\gamma : G\times G \to K$. The function $\gamma$ satisfies the 2-cocycle condition and thus specifies an element of the cohomology group $H^2(G,K)$, which characterizes the central extensions in \eqref{extension}.}

We have just seen how a zero-form UV symmetry $G^{(0)}$ can be transmuted into a one-form IR symmetry $K^{(1)}$.  Motivated by our examples above, it is easy to see that every pair $(G,\ K)$ with \eqref{extension} can in fact be realized in a concrete QFT as symmetries $G^{(0)}$ in the UV and $K^{(1)}$ in the IR. We start with a UV theory based on a $K$ gauge theory with scalar matter fields transforming in an irreducible and faithful representation of the central extension $\tilde{G}$ of $G$ by $K$. (We consider scalar matter fields to avoid any ABJ anomaly.) In the phase where the matter fields are massive, we have this symmetry transmutation at energy scales below the mass of the matter fields.  Note that the actual IR  one-form symmetry can be larger than $K^{(1)}$.

\subsubsection{A simple example}\label{A simple example}

As an example, consider the case of $G = K = \bZ_2$.
We start with a $d+1$-dimensional scalar field theory with global $\bZ_4^{(0)}$ symmetry acting as $\phi \mapsto i \phi$. The scalar potential is
\ie
	V(\phi) = m^2 |\phi|^2 + \mu \phi^4 + \mu^*(\phi^*)^4 +\lambda  |\phi|^4\,, \label{scalar.potential}
	\fe
where, without loss of generality, we can take $\mu$ real and positive.  Next, we gauge the subgroup $\bZ_2 \subset \bZ_4$, i.e., we orbifold by $\bZ_2$.  As in the discussion above, this model has a  $\bZ_4/\bZ_2 \cong \bZ_2^{(0)}$ zero-form global symmetry acting as $\phi \mapsto i \phi$.  It also has a charge conjugation symmetry $\phi \mapsto \phi^*$, which we will ignore, and a
quantum/magnetic $\bZ_2^{(d-1)}$ symmetry due to the gauging.

The UV anomaly is \cite{Tachikawa:2017gyf}
\ie\label{Z4Z2UVa}
	\alpha_\UV = \pi i \int {1\over 2}\delta\tilde{\cal A} \cup \mathcal{C}^{(d)}\,,
\fe
where $\mathcal{A}$ and $\mathcal{C}^{(d)}$ are, respectively, background gauge fields for $\bZ_2^{(0)}$ and $\bZ_2^{(d-1)}$, and ${1\over 2}\delta\tilde{\cal A}$ is obtained by the Bockstein map.

Let us discuss the UV symmetry operators in more detail.  The Wilson line of the $\bZ_2$ gauge theory $W$ is the symmetry operator of the magnetic $\bZ_2^{(d-1)}$ symmetry.   This line can start at $\phi$ and end on $\phi^*$.  Therefore, even though this is a $\bZ_2$ Wilson line, it is equipped with an orientation.  In a way, $W$ is not the same as $W^{-1}$.  Related to that, the zero-form $\bZ_2^{(0)}$ global symmetry acts on it projectively. Depending on its orientation, it acts as $+i$ or $-i$.  This fact reflects the anomaly \eqref{Z4Z2UVa}.

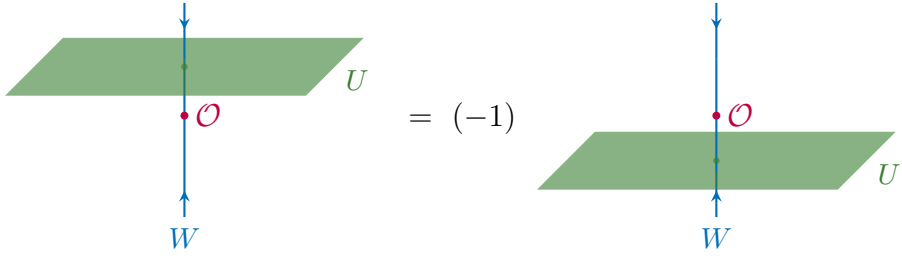
\begin{figure}[ht]
    \centering
    \raisebox{-50pt}{\begin{tikzpicture}[scale=1]
	
	\draw[thick, NavyBlue] (0,-1.6,-1) -- (0,-1,-1);
        \draw[thick, NavyBlue] (0,-1,-1) -- (0,0.5,-1);

        \node[below] at (0,-1.6,-1) {\color{NavyBlue} \small $W$};
        \fill[OliveGreen, opacity=0.6] (-2,0.4,0) -- (2,0.4,0) -- (2,0.4,-2) -- (-2,0.4,-2) -- cycle;
	\draw[thick, NavyBlue] (0,0.5,-1) -- (0,1.25,-1);
	\draw[thick, NavyBlue,-stealth] (0,-1.35,-1) -- (0,-1.25,-1);
        \draw[thick, NavyBlue,-stealth] (0,1,-1) -- (0,0.9,-1);
	
        \fill[OliveGreen, opacity=0.6] (0,0.4,-1) circle (1.25pt);
         \fill[purple] (0,-0.25,-1) circle (1.5pt);
          \node[right] at (0,-0.25,-1) {\color{purple} ${\cal O}$};

        \node[below right] at (2,0.5,-1) {\small \color{OliveGreen} $U$};
    \end{tikzpicture}} $~=~ (-1) ~$
     \raisebox{-50pt}{\begin{tikzpicture}[scale=1]
	
	\draw[thick, NavyBlue,-stealth] (0,-1.35,-1) -- (0,-1.25,-1);
        \draw[thick, NavyBlue,-stealth] (0,1,-1) -- (0,0.9,-1);
	\draw[thick, NavyBlue] (0,-1.6,-1) -- (0,-1,-1);
        \fill[OliveGreen, opacity=0.6] (-2,-0.85,0) -- (2,-0.85,0) -- (2,-0.85,-2) -- (-2,-0.85,-2) -- cycle;
        \draw[thick, NavyBlue] (0,-1,-1) -- (0,0.5,-1);
        \node[below] at (0,-1.6,-1) {\color{NavyBlue} \small $W$};
	\draw[thick, NavyBlue] (0,0.5,-1) -- (0,1.25,-1);

        \fill[OliveGreen, opacity=0.6] (0,-0.85,-1) circle (1.25pt);
         \fill[purple] (0,-0.25,-1) circle (1.5pt);
          \node[right] at (0,-0.25,-1) {\color{purple} ${\cal O}$};

        \node[below right] at (2,-0.75,-1) {\small \color{OliveGreen} $U$};
    \end{tikzpicture}}
    \caption{The topological junction ${\cal O}$ between the ($d-1$)-form symmetry operators/defects $W$ and $W^{-1}$. The point operator ${\cal O}=\phi^2$ is charged under the zero-form symmetry operator $U$, i.e., $U{\cal O} = - {\cal O}U$.}
    \label{fig:2-form}
\end{figure}

Now, consider the local gauge invariant operator ${\cal O}=\phi^2$.  It is odd under the zero-form $\bZ_2^{(0)}$ global symmetry. When $\cal O$ is moved from a generic point in spacetime to the Wilson line $W$, it has the effect of reversing its orientation.  One way to see that is that it changes the eigenvalue of the $\bZ_2^{(0)}$ symmetry operator $U$ from $i$ to $-i$, or the other way around.  (See Figure \ref{fig:2-form}.)  This fact will be important below.

Next, we discuss two IR phases of this model.
\begin{itemize}
\item For $m^2 > 0$, the scalars are massive and the low-energy theory is a pure  $\bZ_2$ gauge theory.  In this phase the UV $\bZ_2^{(0)}$ symmetry is transmuted into an electric $\bZ_2^{(1)}$ one-form symmetry with gauge field ${\cal B}_\IR={1\over 2}\delta\tilde{\cal A}$ and the anomaly \eqref{Z4Z2UVa} matches the IR anomaly
\ie
	\alpha_\IR = \pi i \int  \mathcal{B}_\IR \cup \mathcal{C}^{(d)}\,.
\fe
\item For $m^2 < 0$, the $\bZ_2$ gauge theory is Higgsed, and the UV $\bZ_2^{(0)}$ symmetry is spontaneously broken.  Correspondingly, the low-energy theory has two states.  The order parameter of the spontaneous symmetry breaking is ${\cal O}=\phi^2\sim (\phi^*)^2$.  It becomes the symmetry operator of an IR $d$-form symmetry $\bZ_2^{(d)}$.  We denote its background gauge field by ${\cal D}^{(d+1)}_\IR$.  This symmetry is a result of symmetry transmutation from the UV magnetic $\bZ_2^{(d-1)}$ symmetry via the map ${\cal D}_\IR^{(d+1)}={1\over 2} \delta \tilde {\cal C}^{(d)}$ and its mixed anomaly with the spontaneously broken $\bZ_2^{(0)}$ symmetry
\ie
	\alpha_\IR = \pi i \int  \mathcal{A} \cup \mathcal{D}_\IR^{(d+1)} \,,
\fe
which matches the UV anomaly \eqref{Z4Z2UVa}.  The geometric picture of this symmetry transmutation map is depicted in Figure \ref{fig:2-formUVIR}.  The UV magnetic $\bZ_2^{(d-1)}$ symmetry associated with the background gauge field $\mathcal{C}^{(d)}$, is generated by the topological Wilson line operator, $W$.  In the IR, the Wilson line $W$ does not act faithfully.  However, the point operator $\cal O$ in the junction remains and it acts as the symmetry operator of the new $\bZ_2^{(d)}$ symmetry.
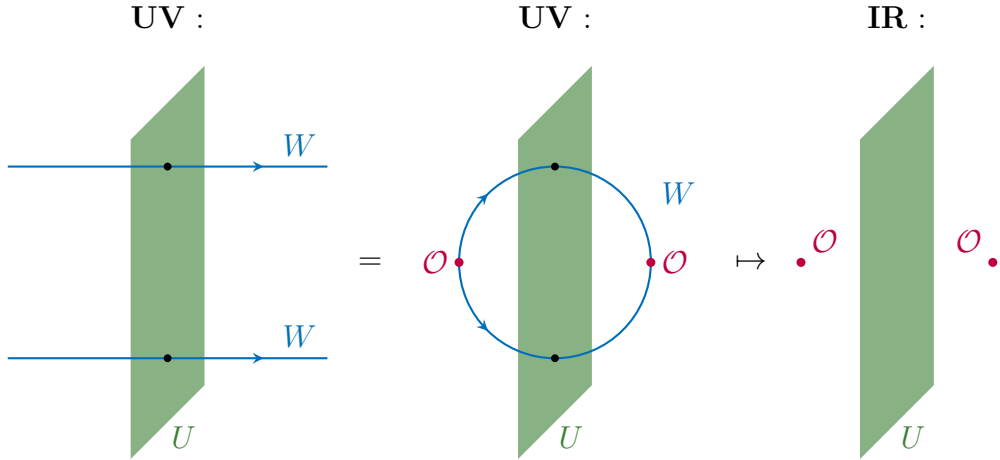
\begin{figure}[ht]
    \centering
    \raisebox{-72pt}{\begin{tikzpicture}[scale=0.85]

     \node[above] at (0,3.5) {\textbf{UV} :};

    \fill[OliveGreen, opacity=0.6] (0,-2.5,1.5) -- (0,2.5,1.5) -- (0,2.5,-1.5) -- (0,-2.5,-1.5) -- cycle;
    \node[below] at (0.25,-2.4) {\color{OliveGreen} $U$};

    \draw[thick, NavyBlue] (-2.5,1.5) -- (2.5,1.5);
    \draw[thick, NavyBlue, -stealth] (1.4,1.5) -- (1.5,1.5);
    \node[above left] at (2.5,1.5) { \color{NavyBlue} $W$};
    \draw[thick, NavyBlue] (-2.5,-1.5) -- (2.5,-1.5);
    \draw[thick, NavyBlue, -stealth] (1.4,-1.5) -- (1.5,-1.5);
    \node[above left] at (2.5,-1.5) { \color{NavyBlue} $W$};

    \fill[black] (0,1.5) circle (1.75pt);
    \fill[black] (0,-1.5) circle (1.75pt);

\end{tikzpicture}} $ ~ = ~ $
	\raisebox{-72pt}{\begin{tikzpicture}[scale=0.85]
	\node[above] at (0,3.5) {\textbf{UV} :};

    \fill[OliveGreen, opacity=0.6] (0,-2.5,1.5) -- (0,2.5,1.5) -- (0,2.5,-1.5) -- (0,-2.5,-1.5) -- cycle;
    \node[below] at (0.25,-2.4) {\color{OliveGreen} $U$};

     \draw[thick, NavyBlue, -stealth] (-1.06,1.06) -- (-1.05,1.07);
     \draw[thick, NavyBlue, -stealth] (-1.06,-1.06) -- (-1.05,-1.07);

    \draw[thick, color=NavyBlue] (0,0) circle (1.5);
    \node[above right] at (1.5,0.75) { \color{NavyBlue} $W$};

    \fill[black] (0,1.5) circle (1.75pt);
    \fill[black] (0,-1.5) circle (1.75pt);

    \fill[purple] (1.5,0) circle (2pt);  \node[right] at (1.5,0) { \color{purple} ${\cal O}$};
    \fill[purple] (-1.5,0) circle (2pt); \node[left] at (-1.5,0) { \color{purple} ${\cal O}$};
\end{tikzpicture}} $~ \mapsto ~$ \raisebox{-72pt}{\begin{tikzpicture}[scale=0.85]
	
    \node[above] at (0,3.5) {\textbf{IR} :};
	
    \fill[OliveGreen, opacity=0.6] (0,-2.5,1.5) -- (0,2.5,1.5) -- (0,2.5,-1.5) -- (0,-2.5,-1.5) -- cycle;
    \node[below] at (0.25,-2.4) {\color{OliveGreen} $U$};

    \fill[purple] (1.5,0) circle (2 pt);  \node[above left] at (1.5,0) { \color{purple} ${\cal O}$};
    \fill[purple] (-1.5,0) circle (2pt); \node[above right] at (-1.5,0) { \color{purple} ${\cal O}$};
\end{tikzpicture}}
\caption{A geometric illustration of the symmetry transmutation from the UV $\bZ_2^{(d-1)}$ into the IR $\bZ_2^{(d)}$ in $d$ spatial dimensions.  It is described by the transmutation map ${\cal D}_\IR^{(d+1)}={1\over 2} \delta \tilde {\cal C}^{(d)}$. On the left, the Wilson line operator, which generates the $\bZ_2^{(d-1)}$ symmetry, acts projectively on the Hilbert space of the $d$-spacetime-dimensional $\bZ_2^{(0)}$ symmetry defect $U$. This configuration is equivalent to the middle diagram. In the IR (rightmost configuration), only the topological junction operators ${\cal O}$ remain. The point operators ${\cal O}$ act on the defect Hilbert space and generate the $\mathbb{Z}_2^{(d)}$ symmetry, which is transmuted from the UV.}
    \label{fig:2-formUVIR}
\end{figure}
\end{itemize}

\subsubsection{Gauge theory examples with gauge/flavor relation}

As we discussed in Section \ref{gaugetheoryexamples}, and demonstrated in the various examples above, including the simple example in Section \ref{A simple example}, symmetry transmutation is generic in gauge theories with massive matter fields that transform in a projective representation of the global symmetry group. In these cases, we study a gauge theory with gauge group $G_\mathrm{gauge}$ and matter fields that transform in an irreducible and faithful representation of
\ie
	\frac{ G_\mathrm{gauge} \rtimes G_\mathrm{naive} }{\cal Z } \,. \label{total.group}
\fe
Here, $G_\mathrm{naive}$ is the naive flavor symmetry group acting on the matter fields, and ${\cal Z }$ is a  subgroup of the center of gauge group $Z(G_\mathrm{gauge})$ and the center of the naive global symmetry $Z(G_\mathrm{naive})$. The global symmetry group that acts faithfully on the gauge-invariant operators constructed out of the fundamental fields is the quotient
\ie
	G = \frac{ G_\mathrm{naive} }{{\cal Z } } \,.
\fe
As in some of the examples above, we neglect possible magnetic symmetries that arise from the gauging.

In the phase where the matter fields are massive, the effective theory below their mass scale is described by a pure gauge theory with gauge group $G_\mathrm{gauge}$.  The center of the gauge group $Z(G_\mathrm{gauge})$ leads to an electric one-form symmetry $Z(G_\mathrm{gauge})^{(1)}$. A part of this one-form symmetry, ${\cal Z}^{(1)}$, is a result of a transmutation from the UV zero-form symmetry $G^{(0)}$.

This happens when $G_\mathrm{naive}$ is a nontrivial extension of $G$ that is characterized by $\Phi \in H^2(G , {\cal Z})$.\footnote{When $G$ is Abelian, $\Phi$ is known as the Bockstein class/homomorphism (see, e.g., \cite{hatcher2002algebraic,Kapustin:2014zva,Benini:2018reh}). For $G$ a Lie group and $\cal Z$ its center, $\Phi$ is known as the Brauer or Stiefel-Whitney class of the principal $G$ bundle \cite{Aharony:2013hda}.}
Then, the map between the UV and IR background gauge fields is determined from the central extension
\ie
1 \to {\cal Z} \to G_\mathrm{naive} \to G \to 1  \,,
\fe
which is characterized by $\Phi \in H^2(G,{\cal Z} )$. Since the matter fields transform under the quotient \eqref{total.group}, the gauge bundle $G_\mathrm{gauge}$ can be twisted depending on the UV background gauge field $A_\UV $ of the global symmetry $G$. For example, the quotient \eqref{1+1d.quotient} in the example of scalar QED$_2$ led to the relation \eqref{bundle.relation} between the SO(3)$_\mathrm{f}$ flavor and $\mathrm{U}(1)_\mathrm{EM}$ gauge bundles. In general, twisting of the gauge bundle corresponds to turning on background gauge fields for the one-form symmetry associated with the center of the gauge group. It follows that turning on a background gauge field $A_\UV $ for the UV symmetry $G^{(0)}$ can lead  to a nontrivial background $B_\IR $ for the IR ${\cal Z}^{(1)}  $ one-form symmetry that is given by
\ie
	B_\IR  = \Phi ( A_\UV  ) ~ \in ~  H^2(\mathcal{M} , {\cal Z} ) \,,
\fe
where $\mathcal{M}$ is the spacetime manifold.  Again, the one-form global symmetry of the IR theory can be larger than ${\cal Z}^{(1)}$.

\subsection{Geometric picture of symmetry transmutation \label{sec:geometric}}

Here we demonstrate the symmetry transmutation from a geometric picture of symmetry defects. For simplicity, we consider the symmetry transmutation of a $\bZ^X_2 \times \bZ^Y_2$ zero-form symmetry into a $\bZ_2^{(1)}$ one-form symmetry in 2+1d described by
\ie
	{\cal B}_\IR  = \mathcal{A}_X \cup \mathcal{A}_Y\,. \label{z2z2.z2.map}
\fe
This generalizes the example in Section \ref{sec:lattice} to 2+1d. Here, $\mathcal{A}_X$ and $\mathcal{A}_Y$ are the background gauge fields for the $\bZ_2^X \times \bZ_2^Y$ UV symmetry and ${\cal B}_\IR $ is the background gauge field for the IR one-form symmetry $\bZ_2^{(1)}$. Geometrically, the intersection of $\bZ^X_2$ and $\bZ^Y_2$ symmetry defects becomes a one-form symmetry defect/operator in the low-energy theory. To see this, we take the Poincare dual of \eqref{z2z2.z2.map} to obtain $C = S_X \cap S_Y$, where $S_X = \star \mathcal{A}_X \in H^2(\mathcal{M}_3,\bZ_2)$, $S_Y = \star \mathcal{A}_Y \in H^2(\mathcal{M}_3,\bZ_2)$, and $C = \star {\cal B}_\IR  \in H^1(\mathcal{M}_3,\bZ_2)$ are the loci of the three kinds of symmetry defects.

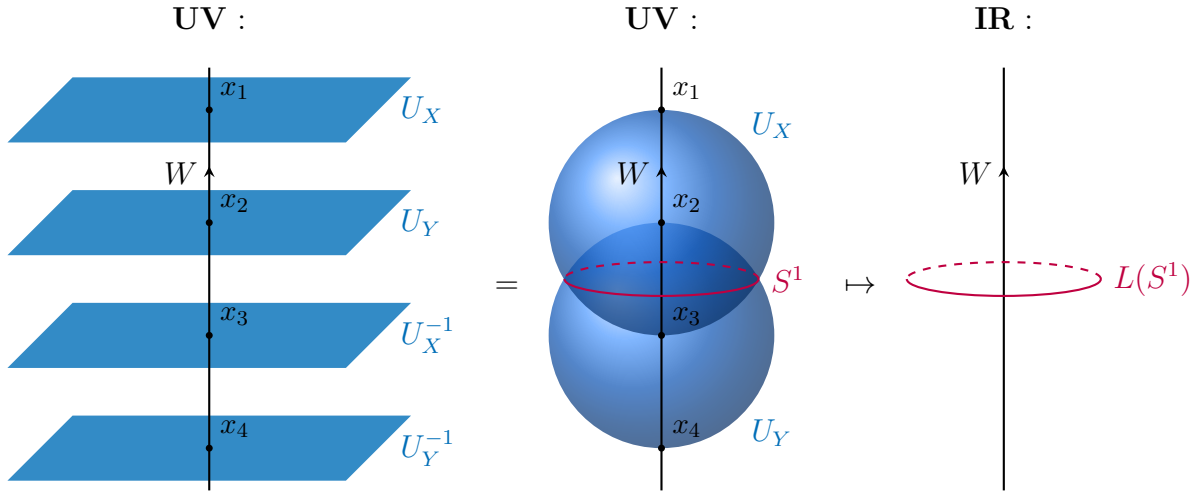
\begin{figure}[th]
    \centering
    \raisebox{-75pt}{\begin{tikzpicture}[scale=0.75]

     \node[above] at (0,4.25) {\textbf{UV} :};
     \fill[NavyBlue, opacity=0.8] (-3,3,1.5) -- (3,3,1.5) -- (3,3,-1.5) -- (-3,3,-1.5) -- cycle;
     \node[right] at (3.2,3) { \color{NavyBlue} $U_X$};

    \fill[NavyBlue, opacity=0.8] (-3,1,1.5) -- (3,1,1.5) -- (3,1,-1.5) -- (-3,1,-1.5) -- cycle;
    \node[right] at (3.2,1) { \color{NavyBlue} $U_Y$};

    \fill[NavyBlue, opacity=0.8] (-3,-1,1.5) -- (3,-1,1.5) -- (3,-1,-1.5) -- (-3,-1,-1.5) -- cycle;
    \node[right] at (3.2,-1) { \color{NavyBlue} $U_X^{-1}$};

    \fill[NavyBlue, opacity=0.8] (-3,-3,1.5) -- (3,-3,1.5) -- (3,-3,-1.5) -- (-3,-3,-1.5) -- cycle;
    \node[right] at (3.2,-3) { \color{NavyBlue} $U_Y^{-1}$};

    \draw[thick, black] (0,-3.75) -- (0,3.75) node[pos=0.75, left] { $W$};
    \draw[thick, black,-stealth] (0,1.9) -- (0,2);

    \fill[black] (0,3) circle (1.75pt); \node[above right] at (0,3) {\small $x_1$};
    \fill[black] (0,1) circle (1.75pt); \node[above right] at (0,1) {\small  $x_2$};
    \fill[black] (0,-1) circle (1.75pt); \node[above right] at (0,-1) {\small  $x_3$};
    \fill[black] (0,-3) circle (1.75pt); \node[above right] at (0,-3) {\small  $x_4$};

\end{tikzpicture}} $~=~$
	\raisebox{-75pt}{\begin{tikzpicture}[scale=0.75]
	
	\node[above] at (0,4.25) {\textbf{UV} :};

    \shade[ball color=NavyBlue, opacity=0.7] (0,-1,0) circle (2);
    \shade[ball color=NavyBlue, opacity=0.7] (0,1,0) circle (2);
    \node[above right] at (1.4,2.3) { \color{NavyBlue} $U_X$};
    \node[below right] at (1.4,-2.3) { \color{NavyBlue} $U_Y$};

    \draw[thick, black] (0,-3.75) -- (0,3.75) node[pos=0.75, left] { $W$};
    \draw[thick, black,-stealth] (0,1.9) -- (0,2);

    \draw[purple,dashed, thick] (1.72,0) arc (0:180:1.72 and 0.3);
    \draw[purple, thick] (-1.72,0) arc (180:360:1.72 and 0.3);
    \node[right] at (1.75,0) {\color{purple} $S^1$};

    \fill[black] (0,3) circle (1.75pt); \node[above right] at (0,3) {\small $x_1$};
    \fill[black] (0,1) circle (1.75pt); \node[above right] at (0,1) {\small  $x_2$};
    \fill[black] (0,-1) circle (1.75pt); \node[above right] at (0,-1) {\small  $x_3$};
    \fill[black] (0,-3) circle (1.75pt); \node[above right] at (0,-3) {\small  $x_4$};
\end{tikzpicture}} $~\mapsto~$ \raisebox{-75pt}{\begin{tikzpicture}[scale=0.75]
	
	\node[above] at (0,4.25) {\textbf{IR} :};

    \draw[thick, black] (0,-3.75) -- (0,3.75) node[pos=0.75, left] { $W$};
    \draw[thick, black,-stealth] (0,1.9) -- (0,2);
)
    \draw[purple,dashed, thick] (1.72,0) arc (0:180:1.72 and 0.3);
    \draw[purple, thick] (-1.72,0) arc (180:360:1.72 and 0.3);
    \node[right] at (1.75,0) {\color{purple} $L(S^1)$};

\end{tikzpicture}}
    \caption{Defects demonstrating the transmutation of a $\bZ_2^X \times \bZ_2^Y$ zero-form symmetry to a $\bZ_2^{(1)}$ one-form symmetry. In the middle panel, the $\bZ_2^X \times \bZ_2^Y$ zero-form symmetry defects $U_X$ and $U_Y$ are inserted along two spheres that intersect each other along the circle $S^1$.  They also intersect the line operator $W$ at the points $x_1,x_2,x_3,x_4$. The zero-form $\bZ_2^X \times \bZ_2^Y$ symmetry could act projectively on $W$, i.e., $U_XU_YU_X^{-1}U_Y^{-1}=\gamma=\pm 1$.  As can be seen in the left panel, the phase $\gamma$ can be determined by shrinking the defects. The right panel configuration describes the IR perspective of it, where the UV zero-form symmetries are transmuted into a $\bZ_2^{(1)}$ one-form symmetry, whose line operator $L(S^1)$ is localized at the intersection circle $S^1$. Shrinking the one-form symmetry line operator should also lead to $\gamma$.  Therefore, $\gamma$ is the result of braiding $W$ and  $L$, i.e., the action of the one-form symmetry on $W$.}
    \label{fig:intersection.picture}
\end{figure}

The geometric picture can be verified by analyzing a line operator $W$ that transforms projectively under the $\bZ_2^X \times \bZ_2^Y$ UV symmetry. Such a line operator is charged under the transmuted one-form symmetry $\bZ_2^{(1)}$. We will show that the intersection of two UV symmetry defects acts nontrivially on such a line operator. To do that, we must show that if the intersection of symmetry defects $U_X$ and $U_Y$ links nontrivially with $W$, then shrinking the intersecting line evaluates to $-1$. This is shown in Figure \ref{fig:intersection.picture}, where we insert symmetry defects $U_X$ and $U_Y$ on two spheres that intersect at a circle $S^1$, which links nontrivially with the Wilson line $W$. In that case, shrinking the defects evaluates to $-1$ because of the projective action of $\bZ_2^X \times \bZ_2^Y$ on $W$.

Another example is the symmetry transmutation of a single $\bZ_2$ zero-form symmetry into a $\bZ_2^{(1)}$ one-form symmetry discussed above around \eqref{scalar.potential}.  Let us specialize to 2+1d and use the map
\ie
	{\cal B}_\IR   =\frac{1}{2} \delta \tilde{\mathcal{A}}\,.
\fe
To consider the geometric picture of this map, we insert the UV symmetry defect on a Klein bottle embedded in the three-dimensional spacetime. The Klein bottle intersects with itself on a circle on which the new one-form symmetry operator of the IR theory is inserted. A similar argument as above shows that this intersection acts nontrivially on line operators that transform projectively under the UV symmetry.

We note that this geometric picture of symmetry transmutation is valid for discrete symmetries. For continuous symmetries, the map between UV and IR background gauge fields, such as $B_\IR  = \mathrm{d} A$ in the fractional quantum Hall, involves non-flat background gauge fields. In some cases, this can be avoided by restricting to a discrete subgroup of the UV symmetry.

\subsection{Emergent anomalies}\label{sec:extension.emergent.anomalies}

In this subsection, we discuss emergent anomalies in higher-form symmetries. Emergent anomalies are anomalies in the IR theory that do not correspond to anomalies in the UV theory \cite{Metlitski:2017fmd, Thorngren:2020wet}. Crucially, the IR symmetry itself is not emergent---only its anomaly is. We will start by reviewing this phenomenon for zero-form symmetries, and then we will extend it to higher-form symmetries and to transmutation.

\subsubsection*{Review of emergent anomalies of zero-form symmetries}

Let us first demonstrate it in an extremely trivial example, consider a QM system with three states with the Hamiltonian $H={\rm diag}(0,0,1)$.  The global symmetry is U(2).  Denoting its representations in an obvious way as ${\bf j}_Q$, the Hilbert space is in ${\bf {1\over 2}}_1\oplus {\bf 0}_0$.  This symmetry is anomaly-free.

The low-energy observer focuses on the two low-lying states in ${\bf {1\over 2}}_1$.  The IR symmetry acting in this low-energy subspace is $\frac{\U(2)}{\U(1)}=\SO(3)$, and it is realized projectively.  The UV to IR homomorphism is
\ie\label{SU2toSO3}
\varphi: \U(2) \to \SO(3)\,.
\fe
Correspondingly, the IR observer thinks that their system has an 't Hooft anomaly explaining why the ground state is nontrivial.

However, as we said above, this anomaly is not present in the UV.  Indeed, we can change the parameters in the Hamiltonian, while preserving its U(2) symmetry and continuously deform it to $H={\rm diag}(1,1,0)$.  Now, the ground state is unique and does not exhibit any symmetry, which is consistent with the assertion that the full system does not have an 't Hooft anomaly.

Importantly, this example differs from many examples with emergent IR symmetries, which have anomalies that do not reflect any anomaly in the UV.  In those more common cases, the relevant IR symmetries are ``new,'' while in our case, the IR symmetry is a quotient of the UV symmetry.

As a higher-dimensional example, consider a 1+1d system with an anomaly-free global $\bZ_4$ symmetry.  This system can have a trivially gapped ground state.  It can also have a non-trivial IR limit, realizing the global $\bZ_4$ symmetry non-faithfully as $\bZ_2$, corresponding to the homomorphism
\ie\label{Z4toZ2}
	\varphi : \bZ_4 \to \bZ_2 \,.
\fe
This low-energy theory can exhibit the $\bZ_2$ anomaly
\ie\label{Z2anomalyIR}
	\alpha[\mathcal{A}_\IR] = \frac{2\pi i}{2} \int \mathcal{A}_\IR \cup \mathcal{A}_\IR \cup \mathcal{A}_\IR \,,
\fe
where $\mathcal{A}_\IR \in H^1(\mathcal{M}, \bZ_2)$ is a $\bZ_2$ gauge field.  As in the QM example above, this anomaly does not exist in the UV  \cite{Thorngren:2020wet}.  To understand how this happens, we should place the UV theory in a background $\bZ_4$ gauge field $\mathcal{A}_\UV$.  Our map $\Phi$ determines the IR gauge field as $\mathcal{A}_\IR=\Phi(\mathcal{A}_\UV)$.  For $\mathcal{A}_\IR$ that arises in this map, the anomaly \eqref{Z2anomalyIR} vanishes, i.e., $\alpha[\Phi(\mathcal{A}_\UV)]$ is a trivial anomaly theory for the $\bZ_4$ gauge field $\mathcal{A}_\UV$.  In other words, the anomaly \eqref{Z2anomalyIR} can be non-zero, but the corresponding gauge fields do not arise as an image of the UV gauge fields $\mathcal{A}_\UV$.

Yet another example is in the context of emanant symmetries. As discussed in \cite{Metlitski:2017fmd,Cheng:2022sgb}, the lattice translation symmetry of the spin-$\frac12$ antiferromagnetic Heisenberg chain leads at low energies to an internal $\bZ_2$ symmetry.  This new $\bZ_2$ IR internal symmetry is not present in the UV.  Yet, it is not an emergent symmetry.  Instead, it is a quotient of the UV translation symmetry and was referred to in \cite{Cheng:2022sgb} as an emanant symmetry (see also \cite{Barkeshli:2025cjs}).  This system has a mixed anomaly between the global SO(3) symmetry and lattice translation, which matches in the IR by a mixed anomaly between the global SO(3) and the emanant $\bZ_2$.  In addition, as in the examples above, the UV lattice translation symmetry is anomaly-free, but the emanant $\bZ_2$ symmetry, which is a quotient of UV lattice translation, can have an emergent anomaly \cite{Metlitski:2017fmd, Thorngren:2020wet}.

More generally, any IR theory with a \emph{0-form} symmetry $G_\IR$ and an 't Hooft anomaly $\alpha[A_\IR]$ can be embedded in a UV system with an appropriate $G_\UV$ global symmetry without that anomaly \cite{Witten:2016cio,Wang:2017loc,Kobayashi:2019lep}.  This means that the onto homomorphism
\ie
\varphi : G_\UV \to G_\IR \,,
\fe
leads to $A_\IR=\Phi(A_\UV)$ such that $\alpha[\Phi(A_\UV)]$ is a trivial anomaly theory for $G_\UV$.\footnote{Mathematically, the pullback of the anomaly with respect to the homomorphism $\varphi$ is trivial.} Here, the map $\Phi$ is derived from the homomorphism $\varphi$ and maps $G_\UV$ gauge fields into $G_\IR$ gauge fields. Note that the homomorphism $\varphi$ needs to be onto, otherwise $G_\IR$, or some part of it, would be emergent. The kinematical statement that the anomaly of $G_\IR$ is trivialized in $G_\UV$ for some homomorphism $\varphi$, is sometimes referred to as \emph{symmetry extension} and was used to construct gapped boundaries of arbitrary SPT phases \cite{Witten:2016cio,Wang:2017loc,Kobayashi:2019lep}.\footnote{Recently, this phenomenon was used to construct ``intrinsically gapless SPTs'' \cite{Thorngren:2020wet,Li:2022jbf,Wen:2022tkg,Antinucci:2024ltv,Ando:2024nlk}, where the IR symmetry $G_\IR$ has an emergent anomaly that forbids the system from being trivially gapped.}

\subsubsection*{Emergent anomalies of higher-form symmetries and transmutation}

Here, we generalize this construction to higher-form symmetries. The novelty here is that the anomaly-free extended UV symmetry must be a lower-form symmetry.  For instance, as shown in \cite{Hsin:2018vcg}, the anomaly of a one-form symmetry in 2+1d is given by the spin of the topological line that generates the one-form symmetry. Therefore, the anomaly cannot vanish by extending the one-form symmetry \emph{group}. However, by finding this symmetry as a result of transmutation of a UV $\U(1)^{(0)}$ (or a discrete subgroup of it) 0-form symmetry, we can trivialize the anomaly as demonstrated in the following.

A known example of this phenomenon was central in the discussion of \cite{NatiSenthilMengAmirSeth}.  There, a gapped Fractional Quantum Hall state with a $\bZ_s^{(1)}$ one-form symmetry with anomaly $r$ was studied.  Coupling the system to  $\bZ_s^{(1)}$ background gauge field ${\cal B}_\IR$, the anomaly theory is
\ie
	\alpha[{\cal B}_\IR] = \frac{2\pi i r}{2s} \int {\cal P}({\cal B}_\IR) \,, \label{zn.oneform.anomaly}
\fe
where $\cal P$ is the Pontryagin square operation.  The anomalous one-form symmetry $\bZ_s^{(1)}$ can originate, via symmetry transmutation from the UV zero-form symmetry U(1)$^{(0)}$.  The map between the UV gauge field $A_\UV$ and the IR gauge field ${\cal B}_\IR$ is
\ie
{\cal B}_\IR=\Phi(A_\UV)=\left[{\mathrm{d}A_\UV\over 2\pi}\right]_s\,.
\fe
For ${\cal B}_\IR$ in the image of this map, the expression \eqref{zn.oneform.anomaly} is nontrivial, but it does not represent an anomaly.  Therefore, the IR anomaly \eqref{zn.oneform.anomaly} is an emergent anomaly.

This example demonstrates that a nonzero anomaly of a $\bZ_s^{(1)}$ one-form symmetry in 2+1d can be an emergent anomaly if this symmetry arises via transmutation from a 0-form symmetry.

\section{Conclusions}\label{Conclusions}

One often defines a global symmetry as associated with an operator that commutes with the Hamiltonian.  For zero-form symmetries, this operator acts on all of space, and for $p$-form symmetries, it acts on co-dimension $p$ in space.  Also, there are additional requirements, like locality.  One can further discuss more exotic symmetries, e.g., non-invertible symmetries, non-topological higher-form symmetries, higher-group symmetries, subsystem symmetries, etc.

Given a global symmetry, it is standard to consider a network of symmetry defects that cannot be localized to a fixed time.  Equivalently, we place the system in spacetime-dependent background gauge fields $A$.  These background gauge fields can be of various forms.

When we analyze the map from the UV theory to the IR theory, the network of defects contains more information than just the symmetries.  In particular, the broader view based on background gauge fields allows us to extend the standard homomorphism between the UV and the IR symmetries \eqref{homomorphism}
\ie
 	\varphi: G_\UV  \to G_\IR  \label{homomorphismc}
\fe
to a homomorphism between UV and IR background gauge fields \eqref{ordinary.map.Phi}
\ie
A_\IR  = \Phi (A_\UV )\,. \label{ordinary.map.Phic}
\fe
In some cases, this is a map from UV background fields of some form to higher-form gauge fields in the IR.  This is the phenomenon of symmetry transmutation.

In this paper, we have explored this phenomenon and emphasized its necessity for matching the symmetries and anomalies between the UV and the IR.

We have emphasized that the homomorphism \eqref{homomorphismc} provides an incomplete picture and should be replaced by the map $\Phi$ between background gauge fields \eqref{ordinary.map.Phic}. For flat background gauge fields, this map is described by a map from networks of symmetry defects in the UV to the networks of symmetry defects in the IR \cite{Barkeshli:2014cna} (see Section \ref{sec:geometric}). Mathematically, the map $\Phi$ between the network of defects can be identified with a map from the \emph{classifying space} of the UV symmetry to the IR higher-form (higher-group) symmetry \cite{Kapustin:2013uxa,Benini:2018reh,Barkeshli:2022edm}. In general, there can be symmetry transmutation from an $n$-group symmetry in the UV into a $p$-group symmetry in the IR with $n \leq p$. Note that symmetry transmutation corresponds to the intersection of symmetry defects that can only increase the codimension of the symmetry defects.

For most of our discussion, we have focused on transmutations of zero-form symmetries in the UV into one-form symmetries in the IR. However, the example in Section \ref{A simple example}, in its broken phase, demonstrated a transmutation of a UV ($d-1$)-form symmetry into an IR $d$-form symmetry.  Then, the analog of the network of UV defects in Figures \ref{fig:intersection.symm.lines} or \ref{fig:intersection.picture} is as in Figure \ref{fig:2-formUVIR}.  Unlike the case of a transmutation of a zero-form symmetry into a one-form symmetry, now the entire network of UV symmetry defects can act at a given time.  Consequently, this network of defects leads to a symmetry operator that acts on the UV Hilbert space at one time.

Even more generally, there can be symmetry transmutation between non-invertible symmetries, which is governed by a map between their corresponding network of defects. For non-invertible symmetries in 1+1d, such a network of defects forms an algebra known as the tube algebra \cite{Lin:2022dhv} (see also \cite{Cordova:2024iti,Choi:2024tri,Gagliano:2025gwr}). Then, the map $\Phi$ should be understood as a map between the tube algebra of the UV and the IR non-invertible symmetries.

Describing the symmetry transmutation map in terms of symmetry defects is particularly advantageous for non-invertible symmetries, where there is no clear notion of background gauge fields. However, this approach has limitations for continuous symmetries with non-flat background gauge fields, as non-flat background gauge fields are not captured by networks of \emph{topological} defects.

Finally, we should stress that our discussion was guided by the simple examples above and those in the context of symmetry fractionalization \cite{Barkeshli:2014cna}. We expect that the full story is considerably richer. First, a more detailed analysis of the map $\Phi$ between UV and IR gauge fields is likely to rely on an interesting mathematical structure, such as the higher homomorphism that was first introduced in the physics literature in \cite{Barkeshli:2014cna} and further explored in  \cite{Barkeshli:2019vtb,Manjunath:2020kne,Aasen:2021vva,Bulmash:2021hmb}. Furthermore, generalizing these conclusions to non-invertible symmetries will phrase this phenomenon in a more categorical language.

\section*{Acknowledgements}

We are grateful to Ofer Aharony, Andrea Antinucci, Maissam Barkeshli, Meng Cheng, Aleksey Cherman, Thomas Dumitrescu, Anton Kapustin, Zohar Komargodski, Shu-Heng Shao, Nikita Sopenko, Yuji Tachikawa, and Ryan Thorngren for interesting discussions. We also thank Maissam Barkeshli, Meng Cheng, Shu-Heng Shao, and Yuji Tachikawa for useful comments on the manuscript. This work was supported in part by DOE grant DE-SC0009988 and the Simons Collaboration on Ultra-Quantum Matter, which is a grant from the Simons Foundation (651444, NS). SS also gratefully acknowledges support from the Sivian Fund and the Paul Dirac Fund at the Institute for Advanced Study. The authors of this paper were ordered alphabetically. This research was supported in part by grant NSF PHY-2309135 to the Kavli Institute for Theoretical Physics (KITP).

\appendix

\section{1-form symmetry anomaly in quantum mechanics \label{app:qm}}

Normally, one does not discuss one-form symmetries in quantum mechanics.  However, by compactifying a higher-dimensional theory on a small space, we effectively find a quantum mechanical system.  And then we could ask about the effect of a one-form symmetry of the higher-dimensional theory in the effective quantum mechanics. Concretely, let us focus on a global U(1)$^{(1)}$ one-form symmetry coupled to a two-form background gauge field $B$.

As we will see in the examples below, the effective quantum mechanical system can have an anomaly theory $ik\int_\Sigma B$. In this case, the Hilbert space of the quantum mechanical model is empty.

Let us discuss it in more detail. On a closed manifold, the gauge transformation $B \mapsto B+d\xi$, with $\xi $ a U(1) gauge field, constrains $k$ to be an integer. On a manifold with a boundary, the partition function of the anomaly theory, $Z=e^{ik\int_\Sigma B} $, is not gauge invariant and transforms as $Z\mapsto Ze^{ik\int_{\partial\Sigma} \xi}$.  To make it gauge invariant, we must add a boundary theory.  The simplest way to do it is to add a pure U(1) gauge theory with a one-form dynamical gauge field $a$ with the boundary action $ik\int_{\partial\Sigma}a$.  Then, if $a$ transforms under $\xi$ as $a\mapsto a-\xi$ the variation of the boundary action cancels the variation of the bulk terms.  However, this boundary theory has no states.  One way to see that is to note that Gauss's law cannot be satisfied for nonzero $k$.  Equivalently, the functional integral over $a$ is an ordinary integral over its holonomy, and it vanishes for nonzero $k$.

When this happens in the effective quantum mechanical problem, it makes the anomaly matching between the UV and IR subtle. Projective phases characterizing the anomaly become ambiguous if the Hilbert space is empty. To address this, one can introduce a non-topological defect in the UV to obtain a non-empty Hilbert space in the IR. See, e.g., the discussion around Figure \ref{circle.with.defect}.

Relatedly, \cite{Delmastro:2022pfo} emphasized a subtlety in the anomaly of ordinary zero-form symmetries in the presence of one-form symmetries with 't Hooft anomalies. Specifically, they highlighted that the anomaly of the zero-form symmetry can change by the field redefinition $B \mapsto B + \rho(A)$, where $A$ and $B$ are the background gauge fields for the zero-form and one-form symmetries, respectively, and $\rho(A)$ is a local map. As mentioned above, this ambiguity in the anomaly of the zero-form symmetry is related to the fact that some Hilbert spaces are empty.  For instance, consider a 0-form symmetry $G^{(0)}$ and a one-form symmetry $K^{(1)}$ with a mixed anomaly between them in 1+1d. The projective action of $G$ on the defect Hilbert space of a $G$ defect is ambiguous since that defect Hilbert space is empty in the IR.

\section{More on the symmetry transmutation map \label{app:math}}

In this Appendix, we make some preliminary comments about the mathematical nature of the symmetry transmutation map $\Phi$ between the UV and IR background gauge fields.

In general, the map $\Phi$ is a map between the background gauge fields of the UV and IR symmetries. (For flat gauge fields, these are networks of defects.)  Here, we focus on a special case and describe this map concretely in terms of the symmetry group in the UV and IR.

We focus on symmetry transmutation from a zero-form symmetry $G^{(0)}$ into a one-form symmetry $K^{(1)}$, and, as always, we denote the groups by $G$ and $K$. Then, the map $\Phi$ has the following  equivalent mathematical presentations:
\begin{enumerate}
	\item A central extension of $G$ by $K$ \label{1}
	\item Homotopy type of a map from $BG$ to $B^2K$ \label{2}
\end{enumerate}

The first description follows from the projective action of $G$ on line operators/defects, as explained in Section \ref{sec:central extension}. Line operators form an Abelian group that is identified with the Pontryagin dual of the one-form symmetry group $K$. Thus, we have a projective action of $G$ that corresponds to a central extension of $G$ by $K$. Such central extensions are classified by the cohomology group $H^2(G,K)$.

The significance of the second point above is that we would like the map $\Phi$ to be independent of the spacetime manifold $\cal M$.  This is needed in order to ensure \emph{locality}.  Then, we use the following crucial property of the classifying space $BG$ of the group $G$: there is a one-to-one correspondence between principal $G$-bundles on a manifold $\mathcal{M}$ and the homotopy type of maps from $\mathcal{M}$ to $BG$. More precisely, any principal $G$-bundle is a pullback of a universal $G$-bundle over $BG$, which is denoted as $EG \to BG$. Similarly, for $B^2K$, there is a bijection between background gauge fields for a one-form symmetry $K^{(1)}$ and the homotopy type of maps from $\mathcal{M}$ to $B^2K$.

The cohomology group $H^2(G, K)$ is identified as the homotopy type of maps from $BG=\mathsf{K}(G,1)$ to $B^2K=\mathsf{K}(K,2)$, which is sometimes denoted by $[BG, B^2 K]$. Such maps are equivalent to maps from $G^{(0)}$ gauge fields to $K^{(1)}$ gauge fields (see, e.g., \cite{hatcher2002algebraic,Kapustin:2013uxa,Tachikawa:2017gyf}). To see this, we first represent a $G^{(0)}$ and $K^{(1)}$ gauge field as maps $A_\UV  : \mathcal{M} \to BG$ and $B_\IR  : \mathcal{M} \to B^2K$. Then,  $\Phi : BG \to B^2K$ maps $A_\UV $ to $B_\IR $ as
\ie
	B_\IR  = \Phi (A_\UV ) ~:~ \mathcal{M} \to B^2 K ~.
\fe

When $G$ is Abelian, and both $G$ and $K$ are discrete, we can be more explicit and expand the discussion of the symmetry transmutation map in Sections \ref{sec:central extension}.  Background gauge fields for $G^{(0)}$ are described by elements of the cohomology group $H^1(\mathcal{M}, G)$, and background gauge fields for $K^{(1)}$ are described by elements of the cohomology group $H^2(\mathcal{M}, K)$. Then, the central extension
\ie
	1 \to K \xrightarrow{i} \tilde{G} \to G \to 1  \label{central.ext}
\fe
leads to the long exact sequence  \cite{hatcher2002algebraic}
\ie
	\cdots \to H^1(\mathcal{M}, K) \to H^1(\mathcal{M}, \tilde{G}) \to H^1(\mathcal{M}, G) \xrightarrow{\Phi} H^2(\mathcal{M}, K) \to \cdots ~,
\fe
and in turn, to the Bockstein homomorphism
\ie
	\Phi( \mathcal{A} ) = i^{-1} \delta \tilde{\mathcal{A}} \,,
\fe	
where $\tilde{\mathcal{A}} \in C^1(\mathcal{M}, \tilde{G})$ is a $\tilde{G}$-lift of the $G$-gauge field $ \mathcal{A} \in Z^1(\mathcal{M}, G)$, and $i^{-1}$ is the inverse of the map $i$ in \eqref{central.ext}. For the case of $G=K=\bZ_2$, we have
\ie
	\Phi(\mathcal{A}) = \frac{1}{2} \delta \tilde{\mathcal{A}} \,.
\fe

Finally, there are symmetry transmutations involving $p$-form symmetries for $p \geq 2$. Patterns of symmetry transmutation from a 0-form symmetry $G^{(0)}$ into a $p$-form symmetry $K^{(p)}$ are described by the cohomology group $H^{p+1}(G, K)$.\footnote{In general, there could be a symmetry transmutation from an $n$-group into a $p$-group for $p \geq n$, which is described by the homotopy type of maps between the classifying spaces of the corresponding higher-groups \cite{Kapustin:2013uxa,Benini:2018reh,Barkeshli:2022edm}.} However, we also expect the existence of \emph{beyond cohomology} symmetry transmutations, i.e., those not captured by the classification $H^{p+1}(G, K)$, which can involve characteristic classes of the spacetime manifold, such as the Stiefel-Whitney classes of the tangent bundle, $\{w_2, w_3, \dots\}$ \cite{Kapustin:2014tfa}. It would be interesting to find gauge theories that realize such symmetry transmutations.

\bibliographystyle{JHEP}

\bibliography{refs}

\end{document}